\documentclass[preprint,12pt]{elsarticle}

\usepackage{amssymb}
\usepackage{amsmath}
\usepackage{graphicx}%
\usepackage{multirow}%
\usepackage{amsmath,amssymb,amsfonts}%
\usepackage{amsthm}%
\usepackage{mathrsfs}%
\usepackage{comment}
\usepackage[title]{appendix}%
\usepackage{xcolor}%
\usepackage{textcomp}%
\usepackage{manyfoot}%
\usepackage{booktabs}%
\usepackage{algorithm}%
\usepackage{algorithmicx}%
\usepackage{algpseudocode}%
\usepackage{listings}%
\usepackage{enumitem}
\usepackage{subcaption}
\usepackage{caption}
\usepackage{hyperref}
\usepackage{natbib}
\usepackage{bbm}
\usepackage{scalefnt}

\definecolor{blue}{HTML}{1F77B4}
\definecolor{orange}{HTML}{FF7F0E}
\definecolor{green}{HTML}{2CA02C}
\definecolor{red}{HTML}{D62728}
\definecolor{purple}{HTML}{9467BD}
\definecolor{brown}{HTML}{8C564B}
\definecolor{pink}{HTML}{E377C2}
\definecolor{grey}{HTML}{7F7F7F}
\definecolor{yellow}{HTML}{BCBD22}
\definecolor{cyan}{HTML}{17BECF}
\definecolor{turquoise}{HTML}{3FE0D0}

\setcitestyle{authoryear,open={(},close={)}} 

\setlist{leftmargin=*, topsep=0.5em, parsep=0pt, itemsep=1em, labelindent=0pt, align=left}

\usepackage{algorithm}
\usepackage{algpseudocode}

\usepackage{longtable}
\journal{arXiv}

\begin{document}

\begin{frontmatter}



\title{Quantformer: from attention to profit with a quantitative transformer trading strategy}


\author[1,3,5]{Zhaofeng Zhang \corref{cor2}} 
\author[1,4,6]{Banghao Chen \corref{cor2}} 
\author[2,1,4]{Shengxin Zhu \corref{cor1}} 
\author[1,3]{Nicolas~Langrené \corref{cor1}} 

\cortext[cor2]{Co-first authors: zhangzf@umich.edu, chenbanghao@u.nus.edu}
\cortext[cor1]{Co-corresponding authors: nicolaslangrene@uic.edu.cn, shengxin.zhu@bnu.edu.cn}

\affiliation[1]{organization={Guangdong Provincial/Zhuhai Key Laboratory of Interdisciplinary Research and Application for Data Science, Beijing Normal-Hong Kong Baptist University},
            city={Zhuhai},
            postcode={519087}, 
            state={Guangdong},
            country={China}}
\affiliation[2]{organization={Advanced Institute of Natural Science, Beijing Normal University},
            city={Zhuhai},
            postcode={519087}, 
            state={Guangdong},
            country={China}}
\affiliation[3]{organization={Department of Mathematical Sciences, Beijing Normal-Hong Kong Baptist University},
            city={Zhuhai},
            postcode={519087}, 
            state={Guangdong},
            country={China}}
\affiliation[4]{organization={Department of Statistics and Data Science, Beijing Normal-Hong Kong Baptist University},
            city={Zhuhai},
            postcode={519087}, 
            state={Guangdong},
            country={China}}
\affiliation[5]{organization={Department of Mathematics, University of Michigan},
            city={Ann~Arbor},
            postcode={48109}, 
            state={Michigan},
            country={United States}}
\affiliation[6]{organization={Department of Biomedical Informatics, Yong Loo Lin School of Medicine, National University of Singapore},
            city={Singapore},
            postcode={119228}, 
            country={Singapore}}

\begin{abstract}
In traditional quantitative trading practice, navigating the complicated and dynamic financial market presents a persistent challenge. Fully capturing various market variables, including long-term information, as well as essential signals that may lead to profit remains a difficult task for learning algorithms. In order to tackle this challenge, this paper introduces quantformer, an enhanced neural network architecture based on transformer, to build investment factors. By transfer learning from sentiment analysis, quantformer not only exploits its original inherent advantages in capturing long-range dependencies and modeling complex data relationships, but is also able to solve tasks with numerical inputs and accurately forecast future returns over a given period. This work collects more than 5,000,000 rolling data of 4,601 stocks in the Chinese capital market from 2010 to 2023. The results of this study demonstrate the model's superior performance in predicting stock trends compared with other 100-factor-based quantitative strategies. Notably, the model's innovative use of transformer-like model to establish factors, in conjunction with market sentiment information, has been shown to enhance the accuracy of trading signals significantly, thereby offering promising implications for the future of quantitative trading strategies. The implementation details and code is available on \href{https://github.com/zhangmordred/QuantFormer}{Github}.
\end{abstract}









\end{frontmatter}


\section{Introduction}
The goal of stock trading is to optimize the return on investment in the capital market according to the process of buying or selling one or more companies' shares. Traders obtain profit when a positive difference is generated by the fluctuation of stock price. However, stocks are influenced by a large number of factors, which constitute a complex system and make it difficult for people to make a profit. The assessment of a stock's evolving trend is inherently challenging due to the highly volatile and interconnected nature of the market, which sets it apart from typical time series modeling \citep{wang2022adaptive}. As a result, many strategies and tools have been built, in parallel with the development of capital markets, and quantitative strategies have been playing an important role among them.

Some traditional quantitative tools, such as the Markowitz portfolio theory \citep{markowitz1952} and the
Capital Asset Pricing Model (CAPM) \citep{Sharpe}, focus mainly on static fundamental analysis. In other words, these strategies aim to make a profit by simple calculation and analysis. Since then, along with the development of computer science, more quantitative methods and tools have been introduced. Within these methods, factor-based strategies have attracted much attention. In 1993, \citet{fama1993common} introduced the Fama-French Three Factor Model (FF3), which has become an influential model in quantitative trading. In 2015, Fama and French revised their model with a Five-Factor Asset Pricing Model (FF5) \citep{fama2015five}. Besides this classical theory, numerous trading strategies have been published for decades. For example, \citet{shi2024dynamic} built an optimal
portfolio to catch Future Investment Opportunities (FIO) by multi-factor models.

\begin{figure}[ht]
\centerline{\includegraphics[width=1.0\textwidth]{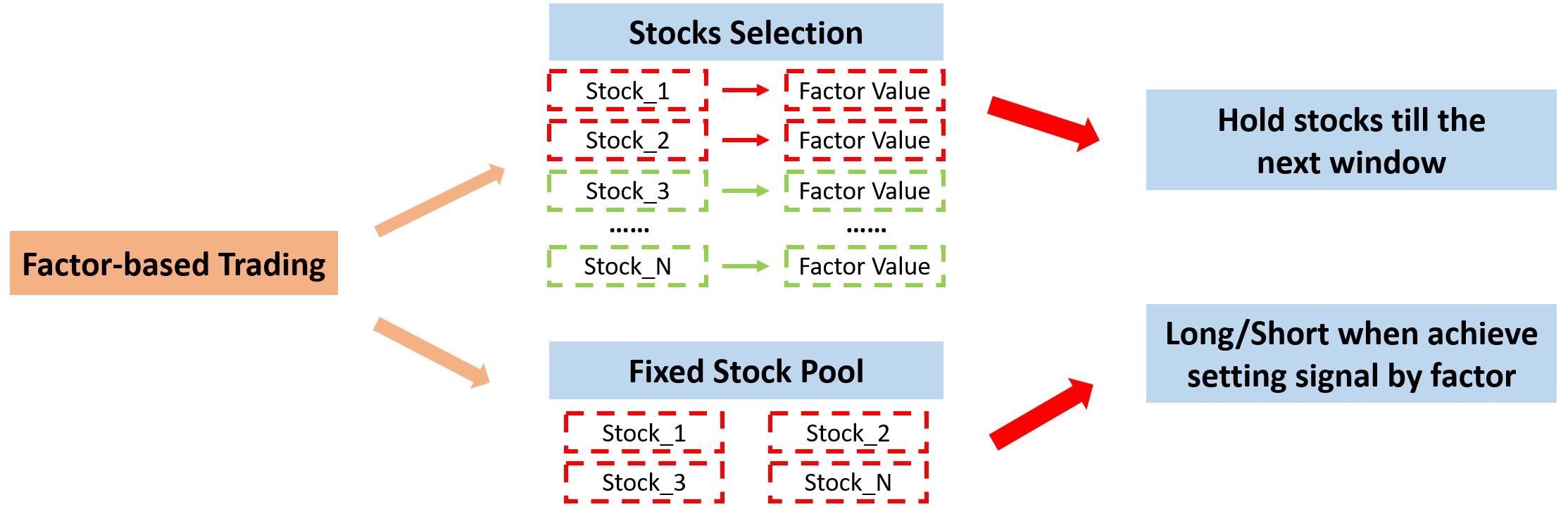}}
\caption{Quantitative trading architecture with factors}
\label{Quantitative Trading1}
\end{figure}

Quantitative trading with factors typically follows two primary approaches, which are shown in Figure \ref{Quantitative Trading1}. The first approach involves the computation of stock factor values. Based on these calculated values, stocks are ranked to establish a pool. Once this pool is established, assets are held for a predetermined period. Adjustments to the portfolio are then made at specific time intervals, ensuring alignment with evolving market conditions and factor readings. The second method employs a fixed pool of stocks, wherein factors guide the derivation of long/short signals. Traders can execute corresponding actions when they receive the signals from factors, allowing for a dynamic response to market fluctuations based on factor insights.

In recent years, Machine Learning (ML) has became an instrumental tool in trading algorithms and decision-making processes. ML methods allow systems to learn from and make decisions based on data, and lend themselves particularly well to the vast and dynamic landscapes of stock markets. For example, \citet{nayak2015naive} exploited machine learning algorithms based on manual indicators, but the underlying random walk hypothesis may hamper the task of understanding inherently non-stationary series. With the rise of different architectures, \citet{feng2019temporal} introduced Relational Stock Ranking (RSR) to catch the sentiment of the market in trading data. 

\begin{figure}[t]
\centering
\includegraphics[width=1.0\textwidth]{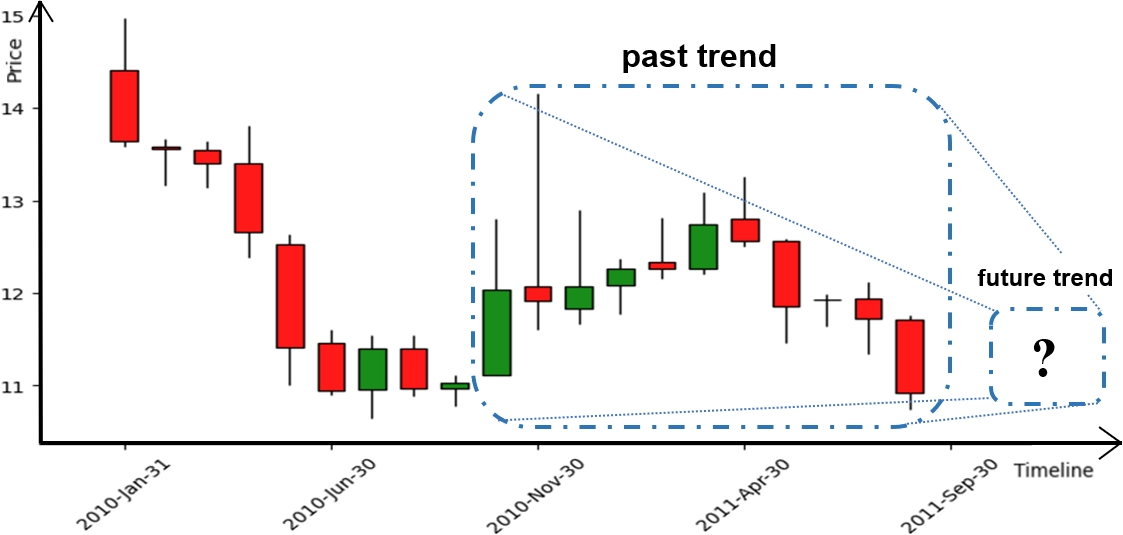}
\renewcommand{\figurename}{Figure}
\caption{Example of stock prediction}
\end{figure}

Although there exist several previous experiments with ML factors that attempted to fetch market sentiment in the quantitative finance field, this research still faces two difficulties. Firstly, in the field of sentiment analysis, which is a branch of Natural Language Processing (NLP, a field of computer science that aims to understand, interpret, and generate human language), models are used to convert words in text to word vectors through word embeddings to serve as inputs. However, instead of words, financial datasets contain both categorical data, such as industry types, as well as quantitative data, such as price fluctuation, turnover rate, and financial indicators. If the input comprises only categorical data, the time series can be treated as a sentence \citep{gorishniy2022embeddings}. In most cases, the input will involve numerical data, which cannot be transformed via word embeddings.

Secondly, most NLP tasks can be transformed into sequence-to-sequence (seq2seq) problems, such as in machine translation, dialogue systems, and speech recognition. As an example, the transformer architecture is based on the seq2seq architecture \citep{10.5555/3295222.3295349}. To utilize existing outputs, decoders in transformer sequentially output samples and use masking operations to handle input sequences during training. However, in stock prediction, where the aim is often to accurately forecast future returns over a period, the transformer model is rarely used for such tasks.

To address these problems, we propose \emph{quantformer}, a modified transformer architecture adapted to quantitative data, and use it as an investment factor. Quantformer is able to input numerical data directly, which refers to a method similar to sentiment analysis. Our contributions lie in following aspects:

\begin{itemize}
    \item We propose quantformer, whose structure adapts to rolling stock-related time series data as inputs without positioning module and word embedding. The new structure with linear embedding shows better fitting to the numerical type inputs.
    \item According to our experiments, quantformer-based factors perform better in back-tests compared with other traditional factor strategies under the same time period with different trading frequencies. These experiments indicate a potential direction for applying transformer-based models to quantitative financial tasks. 
\end{itemize}

The paper is structured as follows. Section~\ref{related} discusses previous quantitative financial works based on machine learning. Sections~\ref{method} and \ref{experiments} introduce quantformers. A factor based on quantformer will be trained and back-tested. For the practical backtest, we collected data from more than 4,600 stocks in the past 14 years (from 2010 to 2023). To comprehensively test the ability of the factor, we divide the data by different frequencies (Subsection~\ref{timestamp}) and trained under different training scales (Subsection~\ref{scales}). Finally, the results of the back-tests including the comparison between the quantformer-factor and other 100 factors as well as the insights gained from such comparative analysis are discussed in Sections~\ref{experiments} and \ref{result}.

\section{Preliminaries} \label{related}
This section briefly introduces the related works about stock prediction with market sentiment and the development of quantitative financial trading methods, especially those based on Machine Learning (ML) methods. 

\subsection{Stock data and market sentiment}
It is known that, in some cases, stock data can be deemed to influence the future trend of a stock price \citep{asness1995power}. \citet{chen2020dynamic, ph2020empirical} have discussed the correlation between investor sentiment and stock market metrics such as turnover rate, indicating that investor feelings, both individual and institutional, significantly influence market behavior, affecting aspects such as volatility and returns. This relationship is not uniform but varies with changes in investor sentiment. It also highlights that investor psychology, especially during periods of high market stress can drastically impact market behavior and investor decisions. This link between sentiment and market performance is further affirmed by studies focusing on factors such as the turnover ratio, which are seen as reflective of market liquidity and investor behavior \citep{naseem2021investor}. \citet{ding2023technical} built a new factor based on sentiment analysis with the average of trading signals from technical trading strategies to benchmark stocks of the S\&P~500 index and DJIA. The sentiment factor shows correlation with stock returns. These findings underscore the mutual relationship between investor sentiment and market performance, demonstrating how psychological factors can drive market dynamics. 

\subsection{Applications of ML methods for trading}
Machine learning exploits a range of algorithms and statistical techniques used for tasks such as regression, clustering, and classification. In past decades, there have been methods that have been increasingly applied to stock data due to their ability to process vast amounts of data and make predictions based on them, offering a potential advantage in the financial markets. At the same time, the finance sector, particularly quantitative trading, has started to be aware of the potential of deep learning models for predicting stock movements, portfolio optimization, and risks. 

\subsubsection{Support Vector Machines (SVMs)}
SVMs have been used in quantitative trading, as they can manage high-dimensional spaces and intricacies found in datasets. \citet{kim2003financial} used SVMs to forecast financial time series in 2003, with a focus on the South Korean stock market, exhibiting the robustness of SVM against traditional statistical methods and some other computational techniques. Similarly, \citet{huang2005forecasting} showcased SVMs' prowess in anticipating the movement direction of Japanese stock markets, especially when determining bullish or bearish market sentiments. \citet{cavalcante2016computational} assessed SVMs' adaptability and precision in financial tasks such as portfolio and quantitative trading.

However, SVMs can also show some disadvantages in quantitative trading. In \citet{huck2009pairs}'s work, SVM was underscored for the accuracy of overfitting. When applied to financial datasets, the model's propensity to fit too closely to the training data might lead to reduced generalization capabilities, making the strategy ineffective on unseen data. In high-frequency trading, though the model displayed competency in specific scenarios, researchers pinpointed potential limitations when grappling with certain volatile market dynamics, suggesting the need for adaptations \citep{kercheval2015modelling}.

\subsubsection{Long Short-Term Memory (LSTM)}
LSTM is a neural network architecture designed to capture long-term dependencies in sequential data \citep{hochreiter1997LSTM}. LSTM networks can accentuate and effectively model volatile financial markets, showcasing significant improvements over traditional time series models. \citet{orsel2022comparative} combined LSTM with linear Kalman filter and set experiments on different volatility stocks to build strategies. \citet{bao2017deep} deployed LSTM with stacked autoencoders to forecast financial time series. \citet{FISCHER2018LSTM} tested the efficacy of LSTM for price predictions. They observed LSTMs' prowess in noisy data and their superiority in prediction accuracy over other conventional models. \citet{sezer2018algorithmic} combined LSTM with other architectures and found that these can enhance the feature extraction process, thereby improving prediction accuracy. For predicting stock prices, \citet{zhang2017stock} harnessed the power of LSTM to unearth multi-frequency trading patterns. Their method underlined the adaptability of LSTMs in modeling complex situations inherent in stock prices. 

\subsubsection{Gated Recurrent Units (GRUs)}
GRUs \citep{chungGRU2014} are another neural network architecture which has emerged as a noteworthy method in the quest for accurate financial forecasting. GRUs can capture long-term dependencies in time series data and have seen increasing popularity in the world of quantitative trading. \citet{bao2017deep} introduced a comprehensive deep-learning framework for financial time series, which incorporated GRUs alongside LSTM and stacked autoencoders. Their model, when applied to diverse datasets, performs better than traditional time series models, accentuating the adaptability and robustness of GRUs. \citet{fischer2018deep} centralized on LSTM networks, provided comparative insights into GRUs. However, in their analysis, LSTMs showed a slight edge in prediction accuracy, indicating that while GRUs are powerful, selecting between them and LSTMs may boil down to specific use cases and computational constraints. Beyond that, \citet {pak2018hybrid} presented a model fusing GRUs with Convolutional Neural Networks (CNN) \citet{oshea2015introduction}. The strategy to combine CNNs and GRUs also exhibits advantages for stock price prediction. \citet{sun2024dynamic} introduced the Attention-GRU model which combined attention and GRU to establish a new factor based on CVaR portfolio \citep{zhu2009worst, ban2018machine}. The Attention-GRU model fitted market return, stocks return from 28 Dow Jones Industrial Average index (DJIA) stocks, which achieved better performance than other models in eight metrics such as annual return, standard deviation and information ratio.

\subsection{Transformer in time series prediction}
Transformer \citep{10.5555/3295222.3295349}, which was introduced in 2017, has received considerable attention in the NLP field. According to its special encoder and decoder stacks with self-attention blocks, the transformer shows its advantage in stability, speed, and long-term memory compared with other traditional models. Transformer-based models such as GPT \citep{Radford2019LanguageMA} and BERT \citep{devlin2018bert} have been influential in the NLP field. More recent transformer-based models include ChatGPT \citep{brownGPT3}, FinBERT \citep{araci2019finbert, yang2020finbert} and BloombergGPT \citep{wu2023bloomberggpt}.

Similar to other neural sequence transduction models, transformer also has an encoder-decoder structure. The model takes as input a sequence of context, which could be natural language sentences, and generates output with texts depending on the input, similar to a translation according to the input or a predicted sentence. Transformer consists of stacks of layers, which include self-attention and point-wise operations. These layers are used to connect the encoder and decoder in the architecture. For these encoder and decoder layers, the self-attention method plays an important role in matrix calculation. The linear and softmax functions are used in the dot-product attention module. In the vanilla transformer model, multi-head attention layers are used and the result from each layer is outputted after going through an additional normalization layer. The above-mentioned steps cover most of the progress made with the vanilla transformer. 

Table \ref{transformerints} shows a set of advanced transformer models, with a focus on time series prediction tasks, which are tasks of interest in quantitative finance. The table highlights the modifications made to the original transformer model. These modifications can be categorized into innovations in attention mechanisms, encoding and decoding structure, positioning module, and mixture model.

Some research proposed enhancements of the attention module. For example, Adversarial Sparse Transformer (AST) \citep{wu2020adversarial} introduced sparse attention using the $\alpha$-Entmax function within adversarial training. Autoformer \citep{wu2021autoformer} replaced traditional self-attention with an auto-correlation mechanism that utilizes periodic patterns in frequency-domain time series analysis. Similarly, Informer \citep{zhou2021informer} suggested ProbSparse attention that selects only the most relevant queries and introduces generative-style decoding to handle long-term dependencies. Gated Transformer Networks (GTN) \citep{liu2021gated} used separate attention for channel-wise and temporal-wise interactions for multivariate time series.  Patch Time Series Transformer (PatchTST) \citep{nie2022PatchTST} divided time series into patches to model local dependencies, and Quatformer \citep{chen2022learning} attempted to use rotation-based attention to capture periodic patterns in time series. AirFormer \citep{liang2023airformer} introduced dartboard spatial attention and causal temporal attention in air quality prediction. Rough transformer \citep{morenopino2024rough} facilitated multi-view signature attention and continuous-time path encoding for financial market modeling for complex financial time series prediction.

In another direction, some works tried to make changes to the encoding or decoding of transformer. FEDformer \citep{zhou2022fedformer} splitted time series data into trend and seasonal components with frequency-enhanced modules. To improve computational efficiency, some models, including Multivariate Transformer \citep{zerveas2021transformer} gave up using a decoder and focused solely on an encoder for feature extraction for forecasting tasks. 

As a kind of revision sight, some research has proposed to replace or extend the positioning modules to better accommodate time series data. For instance, HFformer \citep{barez2023exploring} gave up positioning modules, and uses activation functions with a lightweight linear decoder to revise the model. Noting that temporal order is inherently embedded in time series data, iTransformer \citep{liu2024itransformer} introduced a dimension-inverted structure and removes positioning. Transformer Hawkes Process (THP) \citep{zuo2020transformer} put time intervals into the positioning module directly. FX-Spot Transformer \citep{fischer2024fx} used the Time to Vector (Time2Vec) function to handle irregular time intervals.

To overcome the limitations of standalone transformer models, an approach is to integrate them with other frameworks. ConvLSTM \citep{kim2024physics} merged convolutional neural networks (CNNs) and LSTM layers with transformer moduled to simultaneously model spatial and temporal patterns. ProTran \citep{tang2021probabilistic} integrated state-space models (SSMs) with transformers to address non-Markovian dynamics in time series data, providing robust long-term forecasts. These hybrid approaches improved the versatility and domain adaptability of transformer-based architectures.

\begin{table}
\scriptsize
\caption{Advanced transformer-based architectures for time series data}
\centering
\label{transformerints}
{
\renewcommand{\arraystretch}{1}
\begin{tabular}{p{2.6cm} | p{4cm} p{7.5cm}}
\toprule
Reference & Model & Method\\
\citet{wu2020adversarial} & Adversarial Sparse Transformer (AST)  &  Sparse attention with adversarial training \\
\citet{liang2023airformer} & AirFormer  & Dartboard spatial self-attention and causal temporal self-attention for different training stage \\ 
\citet{wang2022adaptive} & ALSP-TF & Local interaction and global interaction between layers\\
\citet{wu2021autoformer} & Autoformer & Auto-correlation instead of attention \\
\citet{wang2023Bi} & BiLSTM-MTRAN-TCN & Combined BiLSTM, Transformer and Temporary Revolution Network (TCN)\\
\citet{kim2024physics} & ConvLSTM & Combined LSTM with transformer\\
\citet{zeng2023transformers} & LTSF-Linear &  DLinear and NLinear, MLP-based with transformer \\ 
\citet{so2019evolved} & Evolved Transformer & Neural Architecture Search (NAS) in transformer \\ 
\citet{zhou2022fedformer} & FEDformer &  Timestamp encoding with frequency enhanced attention\\ 
\citet{liu2021gated} & Gated Transformer Networks (GTN) & Two-tower transformer with channel-wise \& step-wise\\ 
\citet{ding2020hierarchical} & Gaussian transformer & Multi-scale Gaussian prior, orthogonal
regularization and trading gap splitter for self-attention \\ 
\citet{barez2023exploring} & HFformer & Linear decoder and spiking activation\\ 
\citet{zhou2021informer} & Informer & ProbSparse self-attention, self-attention distilling operation and generative style decoder \\  
\citet{liu2024itransformer} & iTransformer & Dimension inversion and deleted positioning \\  
\citet{li2019enhancing} & LogSparse Transformer &  Causal convolution attention and query-key matching\\ 
\citet{eisenach2020mqtransformer} & MQ-Transformer & Positional encoding from event indicators and decoder-encoder attention for context-alignment \\ 
\citet{ramos2021multi} & Multi-Transformer &  Average multi-head and bagging attention and positioning based on time stamps \\  
\citet{zerveas2021transformer} & Multivariate transformer & Deleted decoder and applied learnable positioning module for multivariate time series\\ 
\citet{liu2022nstrans} & Non-stationary Transformer &  Series standardization and De-stationary attention \\ 
\citet{nie2022PatchTST} & Patch Time Series Transformer (PatchTST) & Divided time series into patches, applied channel-independence and instance normalization \\ 
\citet{tang2021probabilistic} & ProTran &  Modeled non-Markovian dynamics by State Space Models (SSMs) to \\ 
\citet{Liu2022PyraformerLP} & Pyraformer & Pyramidal Attention Module (PAM) and intra-scale neighboring connections model \\ 
\citet{chen2022learning} & Quatformer & Learning-to-rotate Attention (LRA) with global memory and trend normalization\\ 
\citet{kitaev2019reformer} & Reformer & Locality-sensitive hashing attention and reversible residual layers \\ 
\citet{morenopino2024rough} & Rough Transformer & Multi-view signature attention and continuous-time paths encoding transformation\\
\citet{shabani2022scaleformer} & Scaleformer & Multi-scale framework with various models (such as FEDformer, Autoformer) \\ 
\citet{lin2021ssdnet} & State Space Decomposition Neural Network (SSDNet) &  Transformer
 with SSMs  \\ 
\citet{ma2025stockformer} & Stockformer & Dual-frequency spatiotemporal encoder for data in different frequency with graph embedding and fusion attention\\
\citet{chowdhury2022tarnet} & TARNet & Attention score for timestamps masking \\ 
\citet{olorunnimbe2024ensemble} & Temporal Fusion Transformer (TFT) & LSTM learning positional encoding \\ 
\citet{fischer2024fx} & Transformer for FX-Spot &  Time2Vec positioning module\\ 
\citet{zuo2020transformer} & Transformer Hawkes process (THP) &  Positional encoding by translating time intervals\\ 
\citet{cirstea2022triformer} & Triformer &  Triangular,variable-specific patch attention \\ 
\bottomrule
\end{tabular}
}
\end{table}

\subsection{Transformer in quantitative trading}
In the quantitative trading domain, a few research articles have investigated the potential of transformers. \citet{wang2022stock} utilized transformer to predict stock market indices (including CSI~300, S\&P~500, Hang Seng Index, and Nikkei~225) and concluded that the model can better catch the rules of stock market dynamics. \citet{ding2020hierarchical} exploited transformer on trading sequences to classify stock price movements. Besides using a traditional transformer, other innovative methods of transformer are worth mentioning. 
\citet{zhou2021informer} optimized the efficiency of time complexity and memory usage of transformer on extremely long time series by informer. 
\citet{zeng2023financial} combined the advantages of CNNs and transformers to model short-term and long-term dependencies in financial time series. They illustrate the merits of this approach on intraday stock price prediction of S\&P~500 constituents. ConvLSTM \citep{kim2024physics} has been trained on stocks from the S\&P~500 between 2004 and 2021, and performs better than other models such as VAR and ARIMA.

Similar to previous work, \citet{wang2022adaptive} introduced Adaptive Long-Short
Pattern Transformer (ALSP-TF). Their model is structurally designed for hierarchical representation and interaction of stock price series at different context scales. With the help of a learnable function, they make self-attention aware of the weighted time intervals between patterns, to adaptively adjust their dependencies beyond similarity matching. In the end, they obtained more than 10\% of annual return on average.

However, the transformer model can also have some disadvantages in practice. \citet{xu2021relation} mentioned that the global self-attention module focuses on point-wise token similarities without contextual insights. As fluctuations of stocks are conditioned on composite signals over manifold periods, lacking pattern-wise interaction hinders the adequate discrimination of stock tendency and is susceptible to noise points. On the other hand, \citet{wu2020transformer} claimed that the basic query-key matching paradigm is position agnostic. Although position embedding is inserted into the sequential inputs, it may not be optimal because of the inability to reveal precise distances.

\section{Methodology} \label{method}
Several researchers have tried to improve transformers for time series and quantitative financial applications and proposed improved models (see time series papers in Table~\ref{transformerints}), but there are still some aspects that have not been considered. For data in time series type, such as stock prices or daily returns, it is reasonable to assume that each data in the series is assigned a ``position" automatically, which is different from sentences. Based on this observation, the positional encoding module on the vanilla model may not be a necessary part. Furthermore, market data are related to the market trend. For example, the return of the stock may include the market sentiment and the turnover rate reflects investor sentiment and trading enthusiasm. Therefore, integrating such market-related features as input can enhance the model's ability to capture underlying market dynamics and improve the accuracy of future trend predictions. To implement these ideas, we propose \emph{quantformer}, an enhanced transformer-based model for quantitative financial trading, and assess its performance on real-world datasets. 

This section introduces the framework of the work and the steps of the establishment of the model, including data processing, quantformer construction and prediction.

\subsection{Framework overview}
\begin{figure}[ht]
        \begin{center}
        \includegraphics[width=1.0\textwidth]{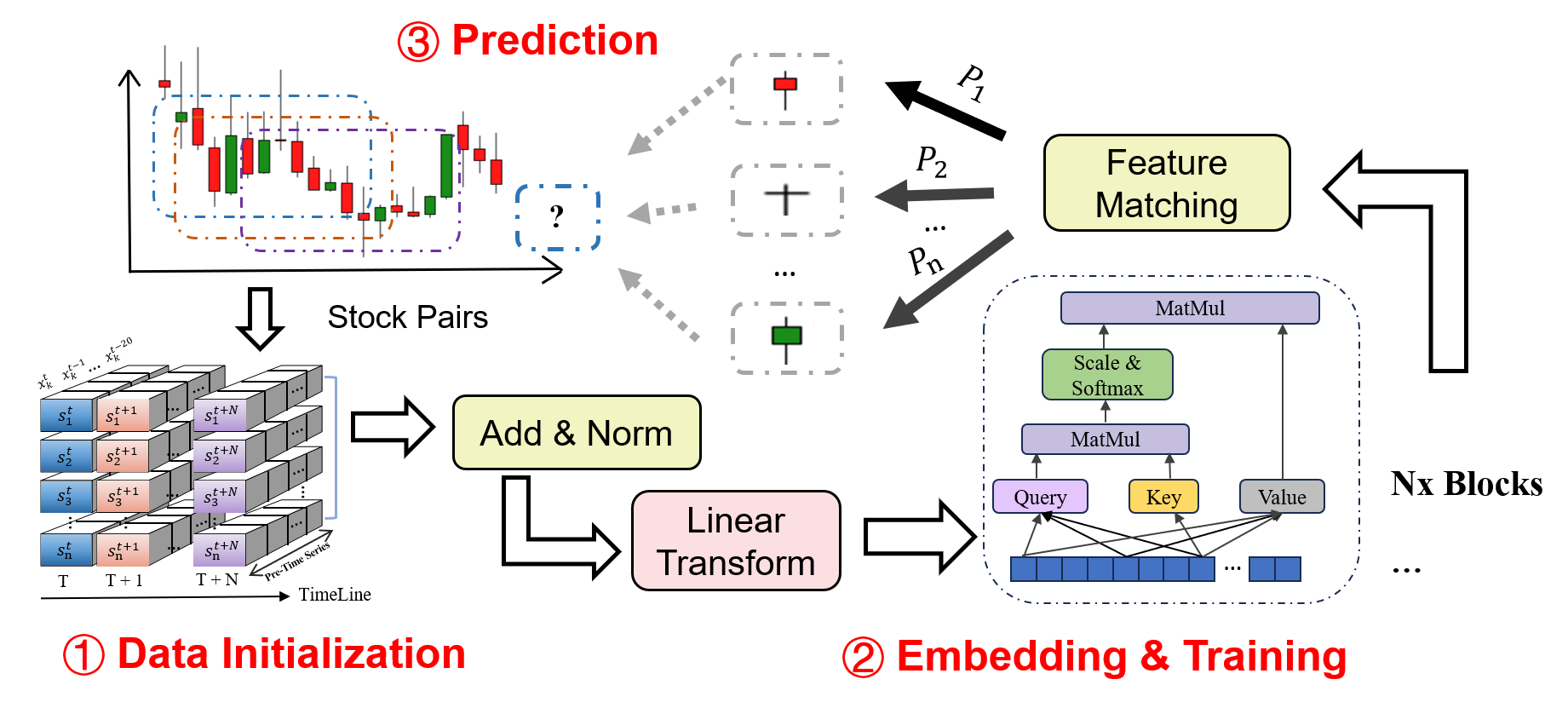}
        \renewcommand{\figurename}{Figure}
        \caption{Overview of the work}
        \label{framework}
        \end{center}
        \end{figure}

Figure~\ref{framework}  gives an overview of the structure. There are three major parts: \\
(i) \textit{Data Initialization} to align the time series of stocks into a regular matrix; \\
(ii) Unlike the transformer, \textit{embedding} the training dataset and \textit{training} them by a quantformer with word-embedding layer replaced by linear layer and with no Mask(opt.) layer;\\
(iii) Lastly, using the trained model to \textit{predict} the possible trend of stocks.

\subsection{Problem formulation}
\label{meth}
\textbf{Training input}\\
Consider a candidate stock set $S^t$ containing $N$ stocks on the trading timestamp $t$: 
$$S^{t} = \lbrace s^{t}_{n} \rbrace^{N}_{n=1}$$
For a stock $s_{n}^{t}$ in the stock set $S^t$, where $n \in \{1,\ldots, N\}$, consider the two-dimensional historical 1-time step (such as one month) input feature matrix $\mathcal{X}^t_n \in \mathbb{R}^{20\times2}$, consisting of 20 continuous timestamps: 
\begin{equation}
{\mathcal{X}}^{t} _{n} = \lbrace x^{t-m}_{n} \vert m = 19, 18, \dots, 0 \rbrace \label{eq:Chi_x}
\end{equation}

Each row vector $x_{n} ^{t-m}$ for $m = 0,\ldots,19$ contains two features. The first one is the accumulated daily profit rate $r^{t-m}_n$ during the time step, where $p^{t-m}_n$ is the close price of the last day at the timestamp $t-m$ and $p^{t-m-1}_n$ is the close price of the last day at the timestamp $t-m-1$. 
\begin{equation}
r^{t-m}_n = \frac{p^{t-m}_n - p^{t-m-1}_n}{p^{t-m-1}_n}, \quad v^{t-m}_n = \sum v^{t-m,i}_{n}
\label{r^{t}}
\end{equation}
The second feature is the accumulated daily turnover rate $v^{t-m}_n$ during the time step $t-m$ for stock $n$. $v^{t-m,i}_n$ is the daily turnover rate and $i$ is the day index in the time period. To sum up, each $\mathcal{X}^{t} _{n}$ can be represented as:
\begin{equation}
\mathcal{X}^{t} _{n} =
\left[
\begin{matrix}
x^{t-19}_n & x^{t-18}_n & \ldots & x^{t}_n \\
\end{matrix}
\right]^T \in \mathbb{R}^{20\times2}, x^{t-m}_n = [r^{t-m}_n, v^{t-m}_n]
\label{eq:Chi_rv}
\end{equation}

\textbf{Normalization}\\
The next step is to normalize the input data for each time step by zero-mean, unit-variance normalization (Z-score). The normalization equation is shown in equation~\eqref{nor}. 
For the row vector $x^t_n$ (equation~\eqref{eq:Chi_x}), $\mathbb{E}[x^{t}]$ and $\mathrm{std}[x^{t}]$ represent the mean and standard deviation of all the row vectors $x^{t}$ at time $t$, respectively, and the resulting values $\tilde{x} ^{t}_{n}$ represent the normalized values of the row data $x^{t}_{n}$.
\begin{equation}
    \tilde{x}^{t}_{n} = \frac{x ^{t}_n - \mathbb{E}[x^{t}]}{\mathrm{std}[x^{t}]}
    \label{nor}
\end{equation}

So, the normalized two-dimensional historical 1-time step input feature matrix can be represented in this way:
\begin{equation}
{\tilde{\mathcal{X}}}^{t} _{n} = \lbrace \tilde{x}^{t-m}_{n} \vert m = 19, 18, \dots, 0 \rbrace 
\label{nor:Chi_x}
\end{equation}

In this way, the input sequences are normalized with zero mean and unit variance, which aims to reduce the influence from outlying time points and allow different features to be comparable with each other \citep{klambauer2017self}.

\textbf{Target labels}\\
For a stock set $S^{t}$ on the trading timestamp $t$, each stock $s^{t}_{n}$ has a profit rate $r_{n}^{t+1}$ on the trading timestamp $t+1$, where the calculation method is the same as for $r^{t}_n$ in the sequence of inputs shown in equation~\eqref{r^{t}}. The set $\mathcal{G}^{t+1}$ contains the next-time stamp's profit for $N$ stocks in the timestamps $t+1$. It is defined as $\mathcal{G}^{t+1} = \{r^{t+1} \}^N_{n=1}$. Each $r^{t+1}_n$ in the list $\mathcal{G}^{t+1}$ is then ranked by $\Psi(r^{t+1}_n)$, which is the empirical quantile CDF for $r^t_n$:

\begin{equation}
\Psi(r^{t+1}_n) = \frac{1}{n} \sum^n_{i=1} \mathbbm{1}_{\{r^{t+1}_i \leq r^{t+1}_n \}}
\label{empirical quantile }
\end{equation}

 For the target label for each feature $s^t_n$ in $\mathcal{G}^{t+1}$, we have a one-hot vector $y^t_n \in \mathbb{R}^{\varrho}$ based on $\Psi(r^{t+1}_n)$. $y^t_n$ is partitioned into $\varrho$ equal bins, where $\varrho$ is set to 3 or more. Each bin corresponds to a $\varphi \times100\%$ of the stocks in the set, where $\varphi\varrho \leq 1$. And the boundary term $\xi$ is used to ensure non-overlapping intervals. Then $y^t_n$ is represented as:
$$
y^t_n=[\mathbbm{1}_{\{ (i-1)(\varrho+\xi) \leq \Psi(r^{t+1}_n) < i\varrho\ + (i-1)\xi\}}  \vert i = 1, \dots, \varrho] \in \mathbb{R}^{\varrho}, \quad \xi = \frac{1-\varphi \varrho}{\varrho-1}
$$
For the empirical quantile CDF $\Psi(r^{t+1}_n)$ of the stock in the range $[i\varphi,(i+1)\xi)$ for $i = 1, \ldots, \varrho$, $y^t_n$ is represented as $\mathbf{0}_{\varrho}$. For example, suppose $\varrho = 3$, $\varphi = 0.2$; for the stocks $s^t_n$ whose profit values $r^t_n$ ranked in the top, middle, bottom $\varphi \times 100\% = 20\%$ respectively, as ($\xi = 0.2$ in this case):
\[
y_n^t =
\begin{cases}
[1, 0, 0]^T & \text{if } \Psi(r_n^{t+1}) \in [0, 0.2) \\
[0, 1, 0]^T & \text{if } \Psi(r_n^{t+1}) \in [0.4, 0.6) \\
[0, 0, 1]^T & \text{if } \Psi(r_n^{t+1}) \in [0.8, 1.0) \\
[0, 0, 0]^T & \text{otherwise}
\end{cases}
\]
The stocks in the intermediate ranges ($\Psi(r^{t+1}_n) \in [0.2, 0.4), [0.6, 0.8)$) are marked as $[0,0,0]^T$. Alternatively, when $\varrho = 5$ and $\varphi =0.2$, the stocks are divided into five equally sized quantiles, with no null labels ($[0,0,0,0,0]^T$), where  each part contains $20\%$ of the stock in $S^t$:
$$y^{t}_{n} \in \{ [1,0,0,0,0]^T, [0,1,0,0,0]^T, \ldots, [0,0,0,0,1]^T\}$$

\subsection{Quantformer encoder}
\label{encoder}
Then, a quantformer encoder structure is designed as described in this subsection. The representation hierarchy consists of $L$ blocks of multi-head self-attention layers. Taking initialized stock embedding sequences $\mathcal{X} = \{\tilde{\mathcal{X}}^{(t)}_n\}^N_{n=1} \in \mathbb{R}^{N\times20\times2}$ as inputs, the canonical self-attention layers introduced in \citet{10.5555/3295222.3295349} can perform information exchange between every time points for each stock $s^t_n$.

To process both categorical and numerical data, the word embedding layer is replaced by a standard linear layer, utilizing linear transformations to substitute the process of word embedding. The linear embedding is shown in equation~\ref{Linear Layer}, where $\mathbf{W}_E \in \mathbb{R}^{2 \times d}$ is the trainable weight matrix and $\theta_E \in \mathbb{R}^d$ is the bias and $d$ is the dimension of the hidden feature space. The resulting sequence $\mathcal{X}'_i$ is
\begin{equation}
\mathcal{X}'_i = \mathcal{X}_i \mathbf{W}_E + \theta_E
\label{Linear Layer}
\end{equation}

In stock prediction, we aim to accurately forecast the return for a future period, thus the model's output is generally a single value representing the probability of price increase or decrease. Therefore, the decoder is simplified by removing the autoregressive prediction mechanism (in the decoder) and the masking operations. For the attention head $h = 1,\ldots , H$, the query, key, and value metrics for each $\mathcal{X}_i \in \mathbb{R}^{20\times2}$ are computed using separate learned linear projections:
\begin{equation}
\mathbf{Q}_{i,h}=\mathcal{X}'_i \mathbf{W}^{Q}_{h}, \quad \mathbf{K}_{i,h}=\mathcal{X}'_i \mathbf{W}^{K}_{h}, \quad \mathbf{V}_{i,h}=\mathcal{X}'_i \mathbf{W}^{V}_{h}
\label{QKV}
\end{equation}
where the trainable weights $\mathbf{W}^{Q}_{h}$, $\mathbf{W}^{Q}_{h}$, $\mathbf{W}^{Q}_{h}$ $\in \mathbb{R}^{d \times \varrho}$ for the $h$-$\mathrm{th}$ head. Then the attention parameter for each head is computed using scaled dot-produced attention: 
\begin{equation}
\mathrm{Attention}(\mathbf{Q}_{i,h}, \mathbf{K}_{i,h}, \mathbf{V}_{i,h}) = \mathrm{softmax}(\mathbf{Q}_{i,h} \mathbf{K}_{i,h}^{T}/\sqrt{d})\mathbf{V}_{i,h}
\label{attention}
\end{equation}
The outputs from all heads are concatenated and sent back to the original dimension using $\mathbf{W}^O \in \mathbb{R}^{\varrho H \times d}$, and the final multi-head attention output $\mathbf{F}_i$:

\begin{equation}
\mathbf{F}_i = \mathrm{Multihead}(\mathbf{Q}_{i,h},\mathbf{K}_{i,h}, \mathbf{V}_{i,h})
=\big(||^{h=H}_{h=1}\mathrm{Attention}(\mathbf{Q}_{i,h}, \mathbf{K}_{i,h}, \mathbf{V}_{i,h})\big)\mathbf{W}^O
\label{multihead}
\end{equation}
where $||$ represents the concatenation operator. After applying attention to all inputs, all the output from attention modules are represented in the form of:
$$\mathbf{F} = [\mathbf{F}_1; \mathbf{F}_2; \ldots \mathbf{F}_N] \in \mathbb{R}^{N\times20\times \varrho H}$$
which are fed into feed-forward layers.

\subsection{Output process}
After processing through the multi-head self-attention and feed-forward layers, with the encoder output $F_i$, the output layer can be represented as:
\begin{equation} 
\mathbf{Z}^t = \mathrm{Softmax}(\mathbf{F}_i \mathbf{W}_Z + \theta_Z)
= \lbrace z^{t}_1, z^{t}_2, \ldots z^{t}_N \rbrace 
\in \mathbb{R}^{\varrho} 
\label{output1} 
\end{equation}
where $\mathbf{W}_Z \in \mathbb{R}^{\varrho \times d}$ and $\theta_Z \in \mathbb{R}^d$. With the softmax function, the predicted probability of the output can be represented as:
    \begin{equation}
        \hat{Y}^{t}_{n} = \{ \hat{y}^t_{n,i} \}^{\varrho}_{i=1} =  \left\{ \frac{\exp(z^t_i)}{\sum^{\varrho}_{i=1} \exp(z^t_i)} \right\}^{\varrho}_{i=1} \in \mathbb{R}^{\varrho}
        \label{output3}
    \end{equation}

In equation~\eqref{output3}, the $\hat{Y}^{t}_{n}$ take values between $0$ and $1$ and sum to $1$, which means that they can be interpreted as the probability distribution of the stock's performance. 

\subsection{Prediction}
The Mean Squared Error Loss (MSELoss) is used to quantify the loss. It is defined as the average of the squares of the differences between the predicted value and the actual prices:
    \begin{equation}
        \mathrm{MSELoss} = \frac{1}{N}  \sum^{N}_{i=1} || (y^t_i - \hat{y}^t_i)||^2_2
        \label{MSE}
    \end{equation}
where $||\cdot||^2_2$ denotes the squared Euclidean distance between predicted and target class probability vectors.

\section{Experiments} \label{experiments}

In this section, we test the ability of the quantformer architecture, introduced in the previous section, for stock price prediction. We are going to detail the set of experiments, including the data resource, implementation details, trading strategy and metrics. 

\subsection{Dataset}
Our training data contains 4601 stocks
listed either on the Shanghai Stock Exchange (SHSE) or on the Shenzhen Stock Exchange (SZSE). The time period ranges from January 2010 to May 2023. The data has been obtained from AKShare\footnote{https://akshare.akfamily.xyz/} and Tushare\footnote{https://www.tushare.pro/}, which are quantitative finance terminals. The training period is from January 2010 to December 2019 and the testing period starts from January 2020.

Closing price adjustments, such as dividends, stock splits, and other corporate actions that can affect a stock's price, are applied to the stock training prices. These adjustments are essential for stock price analysis, especially on long time period, as they provide a more accurate picture of a stock's value and performance over time, which is more appropriate for financial backtesting.
\citep{diamond1971model,wei2022deep}.

\subsection{Time frequency}
\label{timestamp}
Data frequency refers to the number of data points within a specific unit of time, which may reflect different market characteristics \citep{de2018advances}. We consider three frequencies: monthly, weekly, and daily stock data. 

In the first experiments, the time frequency is set to one month. For a stock $s^t_n$ from the stock set $S^{t}$, each item in the feature sequence contains the accumulated profit and accumulated turnover rate in the month (equation~\eqref{eq:Chi_rv}). Sometimes, the available data may not cover whole trading days, for example, because a stock was first listed on the market or resumed trading after an interruption. In such a  situation, we would still record the stock data for this day. However, if there is a whole month during which the stock did not have any trading (perhaps due to stopped trading or not being listed in that month), this month's data will be recorded as ``NaN'', and the sequence $\mathcal{X}^{t} _{n}$ will not be used to train the model.

Within the first experiment, three sub-experiments are set. The sub-experiments share the same inputs. The values $\varrho = 3$ and $\varrho =5$ are used in this experiment. The value of $\varphi$ is fixed for all the experiments to 0.2. In the groups of $\varrho =5$, stocks are divided into five parts and marked respectively. Then, in the groups of $\varrho=3$, the top $20\%$, middle $20\%$ and bottom $20\%$ of the stocks are labeled with one-hot vectors $[1,0,0], [0,1,0]$ and $[0,0,1]$, respectively. Stocks in the intermediate quantile (i.e, $20\%-40\%$ and $60\%-80\%$) are marked as $[0,0,0]$ (as the boundary term $\xi = 0.2$ here, and $\varphi = 0.2$ as well). In the first group of $\varrho=3$, the null labels ($[0,0,0]^T$) are excluded from training, whereas the second group of $\varrho=3$ includes them.

Outliers of both accumulated profit and accumulated turnover rate are not removed from the dataset. These situations may happen in the future and they are expected to be predicted, though these incidents rarely happen.

In the second and third experiments, the inputs are in weekly and daily frequency, respectively. The sub-experiments under different groups are similar to the first one. All the accumulated number of parameters, the number of trained sections (such as 100 months from 2010 to 2019), 
and output dimensions are shown in Table~\ref{para1}.

\begin{table}
\caption{Detailed information about experiments}
\centering
\label{para1}
{
\renewcommand{\arraystretch}{0.5}
\begin{tabular}{l c c c c c}
\toprule
Strategy & Frequency & Lable dim($\varrho$) & Training samples & Section &  Null-label\\ [0.5ex] 
Month\_1 & Monthly & 3 & \enskip 85,490 & 100 & w/o  \\ [0.5ex] 
Month\_2 & Monthly & 3 & 142,409 & 100 & w/ \\ [0.5ex] 
Month\_3 & Monthly & 5 & 142,409 & 100 & w/o    \\ [0.5ex] 
\hline \\
Week\_1 & Weekly & 3 & 455,157 & 466 &  w/o \\ [0.5ex] 
Week\_2 & Weekly & 3 & 758,300 & 466 &  w/ \\ [0.5ex] 
Week\_3 & Weekly & 5 & 758,300 & 466 &  w/o    \\ [0.5ex] 
\hline \\
Day\_1 & Daily & 3 & 3,586,435 & 2,420 & w/o  \\ [0.5ex] 
Day\_2 & Daily & 3 & 5,140,279 & 2,420 & w/ \\ [0.5ex] 
Day\_3 & Daily & 5 & 5,140,279 & 2,420 & w/o  \\ [0.5ex] 
\bottomrule
\end{tabular}
}
\end{table}

\subsection{Implementation details}
The model and its training process are implemented with PyTorch \citep{paszke2019pytorch}. The hyperparameters of the model are optimized by grid search, as it is simple to implement and reliable in low-dimensional spaces \citep{bergstra2012random}. Here the input dimension is 2, and the dimension of the hidden feature space $d$ is 16. The number of multi-head attention modules is 16, and the number of layers of encoder and decoder is 6. The model was trained on an NVIDIA GeForce RTX 2070 GPU and NVIDIA A100 Tensor Core GPU using the Adam optimizer \citep{KingmaB14adam} for 50 epochs. The learning rate is 0.001, and the batch size is 64.

\subsection{Trading strategy}
Algorithm \ref{strategy} shows the pseudocode of the trading strategy. Before the first trade date of the timestamp $t$, all sequences $\mathcal{X}^{t} _{n}$ from the stock set $S^t$ are put in the model and the list of outputs $\hat{Y}^t$ is required. The portfolio value at time $t$, denoted as $P^t$ is updated based on the previous period’s weights and returns as follows:
\begin{equation}
P^t = P^{t-1} (\sum^N_{n=1} w^{t-1}_n (1 + r^{t-1}_n))
\label{valueofstrategy}
\end{equation}
where $w^t_n$ is the weight of stock $s^t_n$,  $r^t_n$ is the return of the stock, and $\sum^N_{n=1} w^t_n = 1$. To determine the weight of each stock, a trading strategy $\Phi \in \mathbb{R}^{\varrho}$ and the decision factor $\mathbf{b}$ is used, where $\mathbf{b}$ determines how many quantile groups are selected, satisfying $ 1 \leq \mathbf{b} < \varrho$. The strategy does not allow short selling (as the Chinese stock market does not allow short selling), so $\Phi$ only contains $0$ or $1$. 
$$
\Phi \in \{0,1\}^{\varrho}, \quad ||\Phi||_0 = \mathbf{b}
$$
For example: 
\begin{itemize}
    \item If $\varrho = 3$, $\mathbf{b} = 1$, then $\boldsymbol{\Phi}$ could be one of: $[1, 0, 0]$, $[0, 1, 0]$, or $[0, 0, 1]$.
    \item If $\mathbf{b} = 2$, then possible values of $\boldsymbol{\Phi}$ include: $[1, 1, 0]$, $[1, 0, 1]$, or $[0, 1, 1]$.
\end{itemize}
Recall the predicted output $\hat{y}^t_n = \{ \hat{y}^t_{n,i} \}^{\varrho}_{i=1}$, we sort $S^t$ by $\Psi(\hat{y}^t_{n,1})$, which is the empirical quantile CDF for $\hat{y}^t_{n,1}$. Similar to the equation~\ref{empirical quantile }, $\Psi(\hat{y}^t_{n,1})$ is computed in this way:
\begin{equation}
\Psi(\hat{y}^t_{n,1}) = \frac{1}{n} \sum^n_{i=1} \mathbbm{1}_{\{\hat{y}^t_{i,1} \leq \hat{y}^t_{n,1} \}}
\label{empirical quantile_y}
\end{equation}
The sorted predicted vector $\tilde{y}^t_n$ for stock $s^t_n$ at time t is shown below. Here the stocks are ranked into $\varrho$ parts and each part contains $\varphi \times100\%$ of the stocks, and the parameters $\varrho$ and $\varphi$ satisfy $\varphi\varrho \leq 1$:
$$
\tilde{y}^t_n=[\mathbbm{1}_{\{ (i-1)(\varrho+\xi) \leq \Psi(\hat{y}^t_{n,1}) < i\varrho\ + (i-1)\xi)}  \vert i = 1, \dots, \varrho] \in \mathbb{R}^{\varrho}, \quad \xi = \frac{1-\varphi \varrho}{\varrho-1}
$$
Then the weight for each selected stock is computed using equation~\ref{weight} shown below, and all the chosen stocks are equal-weighted in this way:
\begin{equation}
w^t_n = \frac{1}{\mathbf{b} \cdot \varrho } (\tilde{y}^{t-1}_n  \Phi^T  \tilde{y}^t_n )^T \cdot \mathbf{1}
\label{weight}
\end{equation}


The same method is run repeatedly during the subsequent periods. The backtest starts from January 2020, in other words, the result of the sequences from May 2018 to December 2019 will be used as the first stock pool to trade. The exchange fee in the Chinese stock market is 0.00384\% (\citealp{sse_charge_table}; \citealp{szse_charge_table}), and the brokerage commission is capped at 0.3\% of the transaction amount (\citealp{transaction_fee_rule}). To keep our backtest results conservative, we set the transaction fee in the backtest to 0.3\%.

\begin{algorithm}[ht]
\caption{Trading strategy}
\begin{algorithmic}[1]
\Require The feature sequence $\chi^{t}_{n}$ for each stock $s^t_n  \in S^t$, initial portfolio cash $P^0$, trading strategy $\Phi$, strategy decision factor $\mathbf{b}$.
\State run\_frequency(trade, frequency, startday=1, time=`open')
\Function{Predict}{$\chi^t_n$}:
    \Return{$\hat{y}^t_n = \text{quantformer}(\chi^t_n)$}
\EndFunction

\Function{SortLabel}{$\hat{y}^t_{n,1}$}:
    \Return{$\tilde{y}^t_n = \left[ \mathbbm{1}\!\left\{\frac{i-1}{\varrho} \leq \Psi(\hat{y}^t_{n,1}) < \frac{i}{\varrho} \right\} \right]_{i=1}^{\varrho}$}
\EndFunction

\Function{ComputeWeight}{$\tilde{y}^{t-1}_n$, $\tilde{y}^t_n$, $\Phi$}:
    \State numerator = $\tilde{y}^{t-1}_n \cdot \Phi^\top \cdot \tilde{y}^t_n$ \;
    \State denominator $= \sum_{k=1}^N \tilde{y}^{t-1}_k \cdot \Phi^\top \cdot \tilde{y}^t_k$ \;
    \State \Return{$w^t_n = \frac{1}{\mathbf{b} \cdot \varrho} \cdot \frac{\text{numerator}}{\text{denominator}}$}
\EndFunction

\Function{Trade}{$w^t_n$, $P^t$}:
    \For{$s^t_n$ \textbf{in} $S^t$}{
        \textbf{order}($s^t_n$, $w^t_n \cdot P^t$)
    }
    \EndFor
\EndFunction
\end{algorithmic}
\label{strategy}
\end{algorithm}
        
\subsection{Metrics}
The Sharpe Ratio (SR) \citep{Sharpe} and the $\alpha$ rate will be used to test the performance of the strategy. The SR is a measure of risk-adjusted return that describes the additional earnings an investor receives for each standard deviation unit increase (equation~\eqref{SR}), where $R_p$ is the return of the portfolio and $R_f$ is the risk-free rate, chosen to be the London Interbank Offered Rate (LIBOR) which averages rate estimates submitted by banks in London\footnote{Although LIBOR officially came to an end on September 30th, 2024, the time period covered by our backtesting dataset is prior to this date. \url{https://www.bankofengland.co.uk/news/2024/october/the-end-of-libor}}.
The risk-free rate in the backtest will be calculated as the average LIBOR rate during the test period.
\begin{equation}
\mathrm{SR} = \frac{\mathbb{E}[R_p]-R_f}{\mathrm{std}[R_p]}
\label{SR}
\end{equation}
        
Alpha represents the excess return of an investment portfolio relative to its benchmark. It aims to measure the stock selection skills of an investment strategy. A positive alpha indicates that the investment portfolio has achieved a higher return than its benchmark after risk adjustment. Specifically, alpha is the excess return of the actual portfolio return over its expected theoretical return. Equation~\eqref{alpha} shows the calculation method of the alpha rate, where $E(R_m)$ is the market return and $\beta$ is the correlation between the portfolio and the benchmark's systemic risk.
\begin{equation}
\alpha = (R_p - R_f)-\beta(\mathbb{E}[R_m]-R_f)
\label{alpha}
\end{equation}
Besides the SR ratio and alpha rate, the annual return (AR) of the stock, the annual excess return (AER), the win rate (WR) and the average turnover rate (TR) of the portfolio will also be shown. The excess returns are returns achieved above and beyond the return of a proxy, chosen to be the CSI~300 index. The win rate is the percentage of trading days during which the portfolio strategy generates a positive return. The turnover rate of the portfolio is the proportion of portfolio weight changes between two consecutive rebalancing periods. The definition of TR at each time $t$ is:
\begin{equation}
\mathrm{TR}^t = \frac{1}{2}\sum^N_{k=1} |w^t_n - w^{t-1}_n |
\label{turnoverp}
\end{equation}
where $w^t_n$ and $w^{t-1}_n$ denote the portfolio weight of the stock $s_n$ in the portfolio at time $t$ and $t-1$, respectively.

Value at risk (VaR) is a method to summarize the total risk in a portfolio \citep{hull2012risk}. 
\begin{equation}
\mathrm{VaR}_{\alpha} = \inf \{x:P(\mathcal{L}>x) \leq \alpha \}
\label{VAR}
\end{equation}

Equation~\eqref{VAR} shows the calculation of VaR, where $\mathcal{L}$ is the loss of the holding period $T$, then $\mathrm{VaR}(\alpha)$ is the $\alpha$-th upper quantile of $\mathcal{L}$. In the measurement of the portfolio, 99$\%$ VaR is used to estimate the maximum loss during the period with 99$\%$ confidence.

\section{Results and discussion} \label{result}
Building upon the analysis of individual strategies, this section presents an evaluation and discussion of the experimental outcomes. The discussion focuses on the comparative effectiveness of different time frequencies (monthly, weekly, and daily) and different training scales.  

\subsection{Overall performance}
Our results are shown in Table~\ref{para}. For the monthly group, Month\_1 shows an annual return of 17.35\% and an annual excess return of 19.43\%, indicating a robust performance over that period. Conversely, Month\_3 shows a diminished annual return and annual excess return of 7.37\% and 9.91\% respectively. The  is notably high in the second month at 51.69\%. The win rate ranges from 43.1\% to 57.3\%, with the first month showing the highest win rate. The Sharpe Ratio and Alpha vary across the strategies, with some periods showing negative values, suggesting underperformance relative to the risk taken, while others show positive values. In the comparison of 99\% VaR, the strategies in monthly frequency perform better than others, where Month\_1 and Month\_3 are lower than 3\%. This means that, with 99\% confidence, the portfolio maximum daily losses are 2.81\% and 2.3\% respectively.

\begin{table}[ht]
\caption{Result of experiments}
\centering
\label{para}
{
\renewcommand{\arraystretch}{0.5}
\begin{tabular}{l c c c c c c c}
\toprule
Strategy & AR & AER & TR & WR & SR & Alpha & VaR\\ [0.5ex] 
Month\_1 & \textbf{17.35\%} & \textbf{19.43\%} & 26.09\% & \textbf{57.8\%} & \textbf{0.915} & \textbf{0.162} & 2.81\\ [0.5ex] 
Month\_2 & \enskip 9.91\% & 13.86\% & \textbf{51.69\%} & 49.3\% & 0.289 & 0.102 & 3.61\\ [0.5ex] 
Month\_3 & \enskip 7.37\% & \enskip 9.91\% & 32.33\% & 51.6\% & 0.246 & 0.064 & \textbf{2.3}\\ [0.5ex] 
\hline \\
Week\_1 & -0.83\% & \enskip 1.31\% & \textbf{7.13\%} & 46.4\% & -0.236 & -0.030 & 3.05\\ [0.5ex] 
Week\_2 & \enskip 7.49\% & 10.81\% & 1.18\% & 49.2\% & 0.160 & 0.085 & 3.73\\ [0.5ex] 
Week\_3 & \textbf{12.3\%} & \textbf{12.73\%} & 1.39\% & \textbf{54.4\%} & \textbf{0.372} & \textbf{0.116} & 3.77\\ [0.5ex] 
\hline \\
Day\_1 & \enskip 7.89\% & \textbf{11.4\%} & \textbf{6.71\%} & 43.1\% & 0.181 & 0.090 & 3.92\\ [0.5ex] 
Day\_2 & \textbf{10.23\%} & 10.94\% & 6.51\% & 44.4\% & \textbf{0.279} & \textbf{0.097} & 3.91\\ [0.5ex] 
Day\_3 & \enskip 9.81\% & 10.03\% & \textbf{5.57\%} & \textbf{44.6\%} & 0.281 & 0.092 & 4.02\\ [0.5ex] 
\hline \\
CSI300 & 1.77\% & \textbackslash & \textbackslash & \textbackslash & -0.015 & \textbackslash & 3.19\% \\ [0.5ex] 
\bottomrule
\end{tabular}
}
\end{table}

In the weekly strategy results, Week\_3 shows an annual return of 12.3\% and an annual excess return of 12.73\%, while Week\_1 displays a negative annual return. For the daily strategies, Day\_2 has a return of 10.23\% and an annual excess return of 10.94\%. The turnover rate across the daily strategies shows lesser variation than the monthly or weekly, indicating a more consistent trading frequency. The win rate for the daily strategies remain fairly stable, hovering around the mid-40\% mark. The Sharpe Ratio and Alpha for weekly and daily strategies show a mix of positive and negative values, reflecting the fluctuating nature of shorter-term trading efficacy.

To demonstrate the advantages of the quantformer factor over traditional factors, 100 price-volume type factors from JoinQuant \footnote{https://www.joinquant.com/} (one of China's largest quantitative finance platforms) are selected for backtesting under the same trading strategy (Algorithm~\ref{Quantitative Trading1}). Price-volume factors are used because these are calculated based on stocks' prices, volumes, and turnover rates, which are similar to the training data used by quantformer factors. The detailed performance of factors and benchmark is shown in \ref{A1}. On average, these factors achieves an annual return of -3.78\%, an average excess return of -2.15\%, and a Sharpe ratio of -0.36, where the benchmark return is -1.77\%. The average maximum drawdown of the strategy is 44.88\%, and the average volatility of the strategy ($\sigma_p$) is 0.26, with the benchmark at 0.197.

\begin{figure}[ht]
\begin{center}
\includegraphics[width=1.0\textwidth]{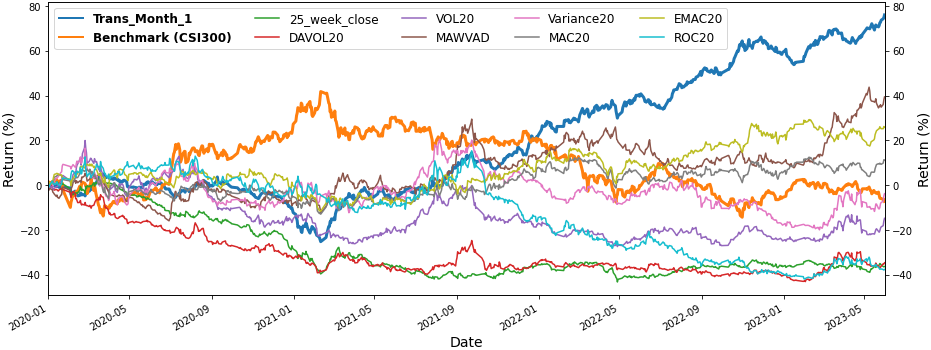}
\renewcommand{\figurename}{Figure}
\caption{Backtest results based on different factors}
\label{backtest_com}
\end{center}
\end{figure}
        
Among these factor-based strategies, the quantformer factor performs well. Except for the win rate, which is slightly lower than a few strategies, it performs well in terms of annual return, annual excess return, $\sigma_p$, Sharpe ratio and Sortino ratio, ranking the best among the 101 factors. The 99\% VaR of the quantformer factor is among the best 10\% factors. Figure~\ref{backtest_com} illustrates the return curves of the quantformer (QF\_Month\_1) alongside the benchmark and some traditional factors. ``25\_week\_rank" represents the current price's position over the past 25 weeks; ``EMAC20" and ``MAC20" are the 20-day index moving average and the stock's 20-day moving average respectively; ``DAVOL20" and ``VOL20" represent the ratio of the 20-day average turnover rate to the 120-day average turnover rate, and the mean of the stock's 20-day turnover rate; ``ROC20" is the price rate of change over 20 days; ``MAWVAD" is calculated as the product of the difference between the closing price and the opening price, divided by the range of the highest and lowest prices, all multiplied by the volume, accumulated over six days; ``Variance20" is the variance of the stock's 20-day annualized returns. Most of these factors are compared based on a window of 20 timestamps, as the quantformer factor was also trained on 20 timestamps. In comparison with these factors, the quantformer factor (blue line) demonstrates a significantly better performance in terms of returns than the benchmark (orange line) and other factor strategies. Compared to other factors by the Sharpe ratio, the quantformer (0.915) still performs better than others (the highest Sharpe ratio among these factors is 0.243). Furthermore, the quantformer also shows a lower downside risk in terms of risk assessment as it achieves a VaR of 2.81, compared to the highest VaR of 2.86 among other factors. These indicate the improvement of the quantformer factor over traditional price-volume factors. 

\subsection{Training under different scales}
\label{scales}
To test the stock selection capabilities of the model, three other factors ``QF\_10\%'', ``QF\_5\%'', ``QF\_1\%'' are trained and tested (with ``QF\_Month\_1\%'', named as ``QF\_20\%''). The four groups are all trained under $\varrho$ = 3, which include the first, middle, and tail 20\%, 10\%, 5\%, and 1\% of stocks in ranking, respectively. The backtest period and other settings are the same as in Section~\ref{experiments}.

\begin{table}[ht]
\caption{Result of factors under different scales}
\centering
\label{scale}
{
\renewcommand{\arraystretch}{0.5}\textbf{}
\begin{tabular}{l c c c c c c c}
\toprule
Strategy & AR & AER & $\sigma_p$ & WR & SR & Alpha & VaR\\ [0.5ex] 
QF\_20\% & 17.35\% & 19.43\% & 0.162 & 57.8\% & 0.915 & 0.162 & 2.81\%\\ [0.5ex] 
QF\_10\% & 13.12\% & 17.73 & 0.159 & \textbf{61.8\%} & 0.574 & 0.128 & 2.023\%  \\ [0.5ex] 
QF\_5\% & 12.59\% & 16.02\%  & \textbf{0.136} & 61\% & 0.63 & 0.117 & \textbf{2.015\%} \\ [0.5ex]
QF\_1\% & \textbf{24.71\%} & \textbf{35.74\%} & 0.214 & 53.3\% & \textbf{0.967} & \textbf{0.249} & 3.048\% \\ [0.5ex]
\hline \\
CSI300 & 1.77\% & \textbackslash & 0.197 & \textbackslash & -0.015 & \textbackslash & 3.19\% \\ [0.5ex] 
\bottomrule
\end{tabular}
}
\end{table}

Table \ref{scale} shows the result of the backtest. QF\_10\% strategy records an AR of 13.12\% and an AER of 17.73\%, with a slightly lower portfolio volatility of 0.159. This strategy has the highest WR at 61.8\%. QF\_5\% strategy delivered 12.59\% AR and 16.02\% AW, with the lowest portfolio volatility at 0.136. It has the lowest VaR at 2.015\%. QF\_1\% strategy outperformed the others, achieving the highest AR of 24.71\% and the highest AER of 35.74\%, though it takes the highest volatility at 0.214. These four factors all perform better than CSI~300 based on their return and risk. These results suggest that the quantformer model's performance seems robust to the choice of different training sets and requirements. 

\section{Conclusion}
This study proposes a new neural network architecture, \emph{quantformer}, inspired by the transformer architecture, for quantitative stock prediction and trading. We address the need for handling numerical input data rather than text, and adapted the model for forecasting tasks rather than sequence-to-sequence problems common in NLP. To enable direct processing of numerical time series data, we replace the word embedding layer with a standard linear layer and removed the output masking operations. We also simplify the decoder to produce a probability distribution over future price movements rather than autoregressively generating token sequences.

We also backtest the quantformer factor on financial market data and compare it with 100 other price-volume factors. Our experimental results demonstrate the promise of this approach. The quantformer-based trading strategies are able to deliver substantial excess returns over the benchmark under a longer time period, with Sharpe Ratios indicating sound risk-adjusted performance. The positive alphas affirm the strategy's ability to outperform expected returns after accounting for risk factors. 

Overall, our work illustrates the viability of the quantformer architecture to handle financial time series, design profitable and robust trading strategies, and calls for further analysis and investigations of this proposed ML-based quantitative methods. The implementation code of quantformer is available at \url{https://github.com/zhangmordred/QuantFormer}.

\section{Acknowledgements and declaration}
This work was funded by the Natural Science Foundation of China (12271047); Guangdong Provincial/Zhuhai Key Laboratory of Interdisciplinary Research and Application for Data Science, Beijing Normal-Hong Kong Baptist University (2022B1212010006); BNBU research grants (UICR0400008-21; UICR0700041-22; R72021114); and the Guangdong College Enhancement and Innovation Program (2021ZDZX1046).

The authors have no competing interests to declare that are relevant to the content of this article.




\bibliography{main}


\appendix
\section{Detailed backtest results between 100 factors and the quantformer factor}\label{A1}
All the data of factors come from JoinQuant\footnote{https://www.joinquant.com/}, which is one of the most common quantitative finance platforms in China. Detailed descriptions of the following factors are available at \url{https://www.joinquant.com/help/api/help#factor_values}. 

\noindent Ten metrics are used to evaluate the performance of each factor:
\begin{itemize}[label={\textbullet}]
  \item Annual Return (\textbf{AR}) and Annual Excess Return (\textbf{AER}) are two metrics to reflect the return of the portfolio directly. The excess returns are returns achieved above and beyond the return of a proxy and the CSI~300 index is used as the basis when calculating excess returns.
  \item Win Rate (\textbf{WR}) is the metric defined as the fraction between the number of trading periods that generate a profit over the total number of trading periods. 
  \item Sharpe Ratio (\textbf{SR}) is a measure of risk-adjusted return that describes the additional earnings an investor receives for each unit of the standard deviation of returns. It is shown in equation~\eqref{SR}. 
  \item The \textbf{Alpha} measures the ability of a portfolio to generate returns above the market benchmark. A positive alpha indicates that the investment portfolio has achieved a higher return than its benchmark after risk adjustment. It is shown in equation~\eqref{alpha}.
  \item \textbf{Beta} is the correlation between the portfolio and systemic risk, reflecting the sensitivity of the strategy to changes in the market. For the daily return of the strategy, $D_p$, and the daily return of the benchmark $D_m$, the equation of beta is:
  $$ \mathrm{Beta} = \beta_p = \frac{\mathbb{C}\mathrm{ov}(D_p, D_m)}{\mathbb{V}\hspace{-0.084em}\mathrm{ar}(D_m)}$$
  where $\mathbb{C}\mathrm{ov}(D_p, D_m)$ is the correlation between between $D_p$ and $D_m$, and $\mathbb{V}\hspace{-0.084em}\mathrm{ar}(D_m)$ is the variance of the daily returns of the benchmark.
  \item Max Drawdown (\textbf{MD}) means the potential worst-case scenario or the most extreme possible loss from the previous peak. For the trough value of a portfolio $A_{\mathrm{trough}}$ and the previous peak value of the portfolio $A_{\mathrm{peak}}$, the MD is calculated in the following way:
   $$ \mathrm{MD} = \frac{A_{\mathrm{peak}} - A_{\mathrm{trough}}}{A_{\mathrm{peak}}}$$
  \item Portfolio volatility ($\sigma_p$) is the standard deviation of the portfolio returns. It provides a kind of weighted average deviation in which large deviations carry more weight \citep{bacon2023practical}. 
  \item Sortino Ratio ($\textbf{STN}$) differentiates harmful volatility from total overall volatility by using the standard deviation of negative portfolio returns. This metric measures the performance of the investment relative to the downward risk.
  \item Value-at-Risk ($\textbf{VaR}$) estimates the potential loss in value of a risky investment, which is used to quantify the amount of potential loss and the likelihood of occurrence for that loss within a specified time frame, which is shown in equation~\eqref{VAR}. In the backtest, 99$\%$VaR is used. 
\end{itemize}
 
{\tiny
\renewcommand{\arraystretch}{1} 
\begin{longtable}{lrrrrrrrrrr}
\toprule               Factor &  AR (\%) &  AER (\%) &  WR (\%) &  SR &  Alpha &  Beta &  MD (\%) &    $\sigma_p$ & VaR (\%) & STN \\
\midrule
\endfirsthead

\toprule
               Factor &  AR (\%) &  AER (\%) &  WR (\%) &  SR &  Alpha &  Beta &  MD (\%) &    $\sigma_p$ & VaR (\%) & STN \\
\midrule
\endhead
\midrule
\multicolumn{1}{l}{{Cont'd}} \\
\midrule
\endfoot

\bottomrule
\endlastfoot
    ARBR &   -1.47 &  0.32 &   54.7 &   -0.251 & -0.002 & 0.914 &     34.46 & 0.221 & 3.74 & -0.375 \\
    AR &   -5.87 &  -4.37 &   53.7 &   -0.406 & -0.048 & 0.981 &     47.84 & 0.260  & 3.94 & 0.612 \\
    ATR14 &   -2.12 &  -0.37 &   62.7 &   -0.250 & 0.006 & 1.160 &     46.55 & 0.250  & 4.51 & -0.35 \\
    ATR6 &   -3.94 &  -2.31 &   61.2 &   -0.328 & -0.014 & 1.175 &     50.11 & 0.253  & 4.48 & -0.455 \\
    BBIC &   1.17 &  3.14 &   50.9 &   -0.109 & 0.033 & 1.038 &     32.64 & 0.258  & 4.16 & -0.155 \\
    BR &   -6.91 &  -5.47 &   53.6 &   -0.467 & -0.064 & 0.934 &     51.72 & 0.254  & 3.80 & -0.718 \\
    CCI10 &   -18.65 &  -17.99 &   43.3 &   -1.402 & -0.261 & 0.812 &     72.04 & 0.220  & 2.79 & -2.026 \\
    CCI15 &   -17.70 &  -16.98 &   44.2 &   -1.293 & -0.240 & 0.832 &     71.43 & 0.223 & 2.90 & -1.866 \\
    CCI20 &   -15.52 &  -14.66 &   46.3 &   -1.051 & -0.196 & 0.868 &     69.12 & 0.235 & 3.28 & -1.519 \\
    CR20 &   -4.27 &  -2.67 &   54.8 &   -0.329 & -0.032 & 0.927 &     49.92 & 0.263 & 4.26 & -0.511 \\
    DAVOL10 &   -7.44 &  -6.04 &   54.1 &   -0.504 & -0.071 & 0.920 &     45.45 & 0.248 & 3.38 & -0.812 \\
    DAVOL20 &   -10.95 &  -9.78 &   54.4 &   -0.702 & -0.119 & 0.919 &     48.96 & 0.246 & 2.86 & -1.079 \\
    DAVOL5 &   -9.54 &  -8.28 &   51.8 &   -0.612 & -0.097 & 0.950 &     46.41 & 0.250 & 3.17 & -1.019 \\
    EMA5 &   2.77 &  4.85 &   48.3 &   -0.047 & 0.049 & 1.041 &     30.70 & 0.259 & 4.66 & -0.068 \\
    EMAC10 &   0.27 &  2.17 &   50.6 &   -0.144 & 0.024 & 1.037 &     31.42 & 0.258 & 4.19 & -0.203 \\
    EMAC12 &   1.58 &  3.57 &   50.6 &   -0.093 & 0.037 & 1.032 &     31.76 & 0.258 & 4.17 & -0.131 \\
    EMAC20 &   6.01 &  8.29 &   54.1 &   0.069 & 0.079 & 1.044 &     32.43 & 0.260 & 4.32 & 0.102 \\
    EMAC26 &   3.83 &  5.97 &   54.1 &   -0.008 & 0.059 & 1.046 &     34.87 & 0.261 & 4.29 & -0.012 \\
    EMAC120 &   1.74 &  3.75 &   48.3 &   -0.088 & 0.034 & 0.960 &     35.81 & 0.254 & 3.77 & -0.145 \\
    Kurtosis20 &   -2.96 &  -1.27 &   52.5 &   -0.395 & -0.027 & 0.772 &     27.71 & 0.182 & 2.74 & -0.6 \\
    Kurtosis60 &   -0.28 &  1.59 &   53.9 &   -0.240 & -0.001 & 0.716 &     22.07 & 0.179 & 2.64 & -0.367 \\
    Kurtosis120 &   -4.12 &  -2.51 &   49.0 &   -0.468 & -0.043 & 0.708 &     28.28 & 0.182 & 2.44 & -0.684 \\
    MAC5 &   0.54 &  2.46 &   46.9 &   -0.134 & 0.026 & 1.035 &     30.03 & 0.257 & 4.29 & -0.195 \\
    MAC10 &   -1.20 &  0.61 &   49.6 &   -0.206 & 0.008 & 1.039 &     34.15 & 0.256 & 4.19 & -0.286 \\
    MAC20 &   2.47 &  4.52 &   53.5 &   -0.059 & 0.046 & 1.043 &     34.58 & 0.257 & 4.09 & -0.083 \\
    MAC60 &   11.10 &  13.72 &   53.8 &   0.243 & 0.120 & 0.987 &     28.61 & 0.254 & 4.34 & 0.399 \\
    MAC120 &   3.11 &  5.21 &   47.7 &   -0.035 & 0.049 & 0.987 &     35.40 & 0.259 & 3.99 & -0.058 \\
    MACDC &   -11.12 &  -9.97 &   52.4 &   -0.674 & -0.117 & 1.003 &     46.48 & 0.261 & 3.54 & -1.046 \\
    MASS &   -10.55 &  -9.35 &   51.0 &   -0.673 & -0.115 & 0.886 &     48.34 & 0.249 & 6.19 & -1.032 \\
    MAWVAD &   7.29 &  9.67 &   55.9 &   0.119 & 0.082 & 0.891 &     36.54 & 0.248 & 4.19 & 0.184 \\
    MFI14 &   -6.21 &  -4.73 &   51.5 &   -0.456 & -0.056 & 0.918 &     45.65 & 0.241 & 3.61 & -0.699 \\
    PLRC6 &   -14.83 &  -13.92 &   47.8 &   -0.905 & -0.178 & 0.960 &     66.24 & 0.259 & 3.85 & -1.368 \\
    PLRC12 &   -7.28 &  -5.87 &   53.3 &   -0.452 & -0.064 & 1.012 &     52.30 & 0.273 & 4.20 & -0.682 \\
    PLRC24 &   -5.77 &  -4.26 &   58.1 &   -0.361 & -0.044 & 1.034 &     53.28 & 0.289 & 4.88 & -0.549 \\
    PSY &   -9.11 &  -7.83 &   50.6 &   -0.660 & -0.096 & 0.880 &     49.79 & 0.223 & 3.36 & -0.992 \\
    Price1M &   -13.55 &  -12.55 &   53.0 &   -0.760 & -0.156 & 0.977 &     64.85 & 0.280 & 4.21 & -1.145 \\
    Price3M &   -6.69 &  -5.25 &   58.4 &   -0.380 & -0.056 & 1.023 &     59.14 & 0.305 & 5.44 & -0.555 \\
    Price1Y &   1.55 &  3.54 &   66.8 &   -0.075 & 0.041 & 1.116 &     60.25 & 0.326 & 7.28 & -0.105 \\
    ROC6 &   -16.13 &  -15.30 &   47.4 &   -0.982 & -0.200 & 0.978 &     68.14 & 0.262 & 3.56 & -1.741 \\
    ROC20 &   -11.63 &  -10.51 &   54.7 &   -0.628 & -0.123 & 1.014 &     60.28 & 0.291 & 4.51 & -0.971 \\
    ROC60 &   -6.86 &  -5.43 &   59.6 &   -0.382 & -0.056 & 1.062 &     62.16 & 0.309 & 5.28 & -0.529 \\
    ROC120 &   2.36 &  4.41 &   66.9 &   -0.051 & 0.049 & 1.117 &  56.36 & 0.319 & 6.79 & -0.071 \\
    Skewness20 &   -5.91 &  -4.41 &   50.8 &   -0.538 & -0.058 & 0.820 &     40.65 & 0.197 & 3.04 & 0.839 \\
    Skewness60 &   -6.47 &  -5.01 &   51.9 &   -0.595 & -0.068 & 0.765 &     38.02 & 0.190 & 2.58 & -0.917  \\
    Skewness120 &   -4.37 &  -2.77 &   51.1 &   -0.459 & 0.044 & 0.743 &     36.01 & 0.192 & 2.69 & -0.684 \\
    TRIX5 &   -10.72 &  -9.54 &   52.5 &   -0.607 & -0.110 & 1.009 &     56.07 & 0.280 & 4.16 & -0.923 \\
    TRIX10 &   -3.51 &  -1.85 &   58.4 &   -0.268 & -0.019 & 1.008 &     52.92 & 0.292 & 5.14 & -0.4  \\
    TVMA20 &   -1.92 &  -0.16 &   64.9 &   -0.209 & 0.014 & 1.268 &     44.66 & 0.288 & 4.63 & -0.31 \\
    TVMA6 &   -0.75 &  1.09 &   62.4 &   -0.166 & 0.026 & 1.259 &     39.46 & 0.288 & 4.76 & -0.246  \\
    TVSTD20 &   -1.83 &  -0.06 &   58.5 &   -0.214 & 0.012 & 1.218 &   35.89 & 0.277 & 4.03 & -0.318 \\
    TVSTD6 &   -3.77 &  -2.13 &   53.4 &   -0.297 & -0.010 & 1.202 &   46.57 & 0.272 & 4.40 & -0.449 \\
    VDEA &   -3.05 &  -1.37 &   57.6 &   -0.305 & -0.023 & 0.857 &     37.01 & 0.239 & 3.06 & -0.484  \\
    VDIFF &   -7.93 &  -6.57 &   0.2 &   -0.533 & -0.079 & 0.895 &     39.79 & 0.247 & 3.08 & -0.842  \\
    VEMA5 &   -0.41 &  1.46 &   57.9 &   -0.186 & 0.006 & 0.850 &     37.31 & 0.237 & 3.33& -0.277 \\
    VEMA10 &   0.95 &  2.91 &   58.9 &   -0.128 & 0.019 & 0.844 &     35.51 & 0.236 & 3.36 &-0.19  \\
    VEMA12 &   1.01 &  2.96 &   59.3 &   -0.126 & 0.020 & 0.840 &     35.17 & 0.235 & 3.35 & -0.186  \\
    VEMA26 &   2.41 &  4.46 &   59.3 &   -0.068 & 0.033 & 0.838 &     31.03 & 0.231 & 3.37& -0.1  \\
    VMACD &   -1.03 &  0.79 &   50.5 &   -0.216 & 0.001 & 0.875 &     33.96 & 0.235 & 3.38 & -0.345 \\
    VOL5 &   -8.12 &  -6.76 &   55.9 &   -0.452 & -0.066 & 1.155 &     53.86 & 0.296 & 4.48 & -0.656  \\
    VOL10 &   -7.07 &  -5.64 &   58.3 &   -0.404 & -0.053 & 1.145 &     50.65 & 0.298 & 4.35 & -0.59 \\
    VOL20 &   -5.43 &  -3.89 &   58.6 &   -0.337 & -0.032 & 1.156 &     50.19 & 0.298 & 4.33&  -0.481 \\
    VOL60 &   0.42 &  2.34 &   58.1 &   -0.125 & 0.032 & 1.145 &     38.35 & 0.286 & 4.71 & -0.175 \\
    VOL120 &   -1.88 &  -0.12 &   53.0 &   -0.213 & 0.007 & 1.136 &     42.09 & 0.281 & 4.64 & -0.3 \\
    VOL240 &   -5.51 &  -3.98 &   47.0 &   -0.382 & -0.038 & 1.075 &     43.53 & 0.265 & 3.99 &  -0.541\\
    VOSC &   -6.35 &  -4.88 &   51.2 &   -0.453 & -0.056 & 0.944 &     40.40 & 0.246 & 3.44 &  -0.739 \\
    VROC6 &   -16.30 &  -15.49 &   42.7 &   -1.179 & -0.208 & 0.897 &     69.03 & 0.221 & 2.91 & -1.778  \\
    VROC12 &   -11.47 &  -10.34 &   46.7 &   -0.793 & -0.127 & 0.915 &     51.62 & 0.228 & 2.74 & -1.241 \\
    VR &   -8.43 &  -7.10 &   52.8 &   -0.566 & -0.083 & 0.932 &     48.57 & 0.244 & 3.72 &  -0.834 \\
    VSTD10 &   -0.77 &  1.07 &   56.3 &   -0.205 & 0.001 & 0.843 &     37.75 & 0.234 & 3.19 &  -0.309 \\
    VSTD20 &   -1.60 &  0.19 &   53.7 &   -0.245 & -0.007 & 0.844 &     37.31 & 0.232 & 3.01 &  -0.366 \\
    Variance20 &   -1.94 &  -0.18 &   59.9 &   -0.201 & 0.005 & 1.124 &     46.38 & 0.302 & 4.84 &  -0.301 \\
    Variance60 &   9.96 &  12.51 &   65.2 &   0.168 & 0.122 & 1.178 &     30.63 & 0.313 & 5.91 & 0.241  \\
    Variance120 &   5.40 &  7.65 &   60.8 &   0.040 & 0.083 & 1.193 &     39.54 & 0.312 & 5.93  & 0.058 \\
    Volume1M &   -15.54 &  -14.68 &   20.7 &   -0.867 & -0.189 & 0.991 &     63.68 & 0.285 & 3.61 &  -1.339\\
    WVAD &   -0.71 &  1.13 &   52.3 &   -0.199 & 0.002 & 0.836 &     42.38 & 0.238 & 3.67 & -0.317 \\
    arron\_down\_25 &   -7.35 &  -5.94 &   48.3 &   -0.593 & -0.072 & 0.880 &     42.63 & 0.209 & 2.84 & -0.826  \\
    arron\_up\_25 &   -15.51 &  -14.65 &   46.4 &   -1.049 & -0.195 & 0.870 &     66.73 & 0.235 & 3.19 & -1.604 \\
    bear\_power &   -13.31 &  -12.30 &   50.8 &   -0.800 & -0.154 & 0.934 &     59.58 & 0.261 & 3.61 & -1.182 \\
    beta &   10.30 &  12.87 &   62.4 &   0.168 & 0.134 & \textbf{1.344} &     35.96 & 0.328 & 6.76 & 0.252 \\
    book\_to\_price &   1.87 &  3.86 &   58.5 &   -0.117 & 0.031 & 0.889 &     26.73 & 0.181 & 2.91 & -0.162 \\
    bull\_power &   -14.81 &  -13.91 &   51.1 &   -0.844 & -0.176 & 0.984 &     65.99 & 0.278  & 3.83 & -1.285 \\
    earnings\_yield &   -2.41 &  -0.68 &   52.6 &   -0.348 & -0.029 & 0.627 &     25.87 & 0.188  & $\textbf{2.55}$ & -0.55\\
    growth &   2.90 &  4.99 &   59.9 &   -0.045 & 0.052 & 1.068 &     37.35 & 0.244  & 4.50 & -0.067\\
    leverage &   2.05 &  4.07 &   53.1 &   -0.089 & 0.023 & 0.719 &     22.57 & 0.217 & 3.22 & -0.144 \\
    liquidity &   -2.71 &  -1.00 &   56.7 &   -0.243 & -0.002 & 1.137 &     41.28 & 0.284  & 4.56 & -0.35\\
    momentum &   10.93 &  13.54 &   65.8 &   0.195 & 0.129 & 1.159 &     43.19 & 0.310 & 6.82 &  0.271\\
    money\_flow\_20 &   -1.72 &  0.05 &   65.5 &   -0.202 & 0.016 & 1.269 &     44.10 & 0.288 & 4.63 & -0.298 \\
    price\_no\_fq &   6.83 &  9.17 &   \textbf{76.1} &   0.095 & 0.092 & 1.132 &     44.86 & 0.267 & 6.20 &  0.131\\
    pull-up &   2.19 &  4.23 &   54.5 &   -0.068 & 0.044 & 1.062 &     33.09 & 0.265 & 3.92 & -0.093 \\
    pull-down &   -2.28 &  -0.54 &   47.6 &   -0.317 & -0.015 & 0.831 &     34.46 & 0.203 & 3.87 & -0.439 \\
    residual\_volatility &   4.13 &  6.30 &   62.3 &   0.003 & 0.062 & 1.048 &     41.34 & 0.269 & 5.25 & 0.004 \\
    sharpe\_ratio\_20 &   -11.88 &  -10.78 &   54.0 &   -0.663 & -0.129 & 0.979 &     57.66 & 0.282  & 4.09 & -1.019\\
    sharpe\_ratio\_60 &   -7.42 &  -6.02 &   58.8 &   -0.417 & -0.065 & 1.029 &     65.19 & 0.300 & 5.17 & -0.583 \\
    sharpe\_ratio\_120 &   -0.32 &  1.55 &   65.7 &   -0.143 & 0.019 & 1.062 &     60.52 & 0.303 & 6.10 & -0.2 \\
    size &   -6.36 &  -4.89 &   51.8 &   -0.571 & -0.058 & 0.906 &     36.00 & 0.196 & 2.64 &  -0.82 \\
    turnover\_volatility &   -10.80 &  -9.62 &   51.8 &   -0.624 & -0.108 & 1.075 &     57.05 & 0.274  & 3.70 & -0.905\\
    1day\_VPT &   -16.45 &  -15.65 &   44.8 &   -1.062 & -0.211 & 0.904 &     70.43 & 0.248 & 3.09 & -1.553 \\
    1day\_VPT\_6 &   -6.36 &  -4.89 &   53.0 &   -0.416 & -0.053 & 0.995 &     47.06 & 0.269 & 3.81 &  -0.82\\
    1day\_VPT\_12 &   -0.07 &  1.82 &   57.9 &   -0.146 & 0.020 & 1.033 &     46.51 & 0.278 & 4.39 &  -0.22 \\
    25week\_close &   -11.51 &  -10.39 &   51.0 &   -0.859 & -0.133 & 0.817 &     47.79 & 0.211 & 2.44 &  -0.684\\
    $\textbf{QF\_Month\_1}$ &   $\textbf{17.35}$ &             $\textbf{19.43}$ &    57.8 &        $ \textbf{0.915}$  &  \textbf{0.162} & 0.588 &   $\textbf{18.35}$ & $\textbf{0.161}$  & 2.81 & $\textbf{0.946}$ \\
    $\textbf{QF\_Month\_2}$ & 9.91  & 13.53     & 49.3   &  0.289   & 0.102  & 0.684 &  26.77 & 0.205  & 3.61 & 0.372 \\
   $\textbf{QF\_Month\_3}$ &  7.37  &  10.4     &  51.6  & 0.246    & 0.064  & 0.483 & 17.33  & 0.137  & 2.3 & 0.309 \\
   $\textbf{QF\_Week\_1}$ &  -0.83  & 1.41     &  46.4  &  -0.236     & -0.003  & 0.726 & 29.31  & 0.204  & 3.05 & -0.291 \\
   $\textbf{QF\_Week\_2}$ &  7.49  & 10.55    & 49.2   & 0.16      & 0.085  & 0.792 & 29.08  & 0.219  & 3.73 & 0.206 \\
   $\textbf{QF\_Week\_3}$ &  12.3  &   12.4   &  54.4  &  0.372     & 0.116  & 0.793 & 29.86  &  0.223 & 3.77 & 0.48 \\
   $\textbf{QF\_Day\_1}$ & 7.89   &  11.4   & 43.1   &  0.181     & 0.09  & 0.834 & 29.15  & 0.196  & 3.92 & 0.245 \\
   $\textbf{QF\_Day\_2}$ &   10.23 &  10.68  & 44.4   & 0.279    &  0.097 & 0.819 & 30.02  & 0.223  & 3.91 & 0.361 \\
   $\textbf{QF\_Day\_3}$ &  9.81  &  10.03     &  44.6  &  0.284     & 0.098  & 0.801 & 30.1  &  0.223 & 1.02 & 0.367 \\
   $\textbf{QF\_10\%}$ &  13.12  &  17.73    & 61.8   & 0.754     & 0.128  & 0.594 & 18.32  & 0.159  & 2.02 & 0.714 \\
    $\textbf{QF\_5\%}$ &  12.59  &  16.02 &  61  &  0.63     & 0.117  & 0.504 & \textbf{13.3}  & \textbf{0.136}  & 2.01 &  0.796\\
    $\textbf{QF\_1\%}$ &  \textbf{24.71}  &  \textbf{35.34}     &  53.3  &  \textbf{0.967}     & \textbf{0.249}  & 0.667 & 23.28  & 0.214  & 3.35 & \textbf{1.299} \\
    Average	& -2.92 & -1.25	& 53.28	& -0.32	& -0.02	& 0.96	& 43.61	& 0.25 & 3.99 & -0.467 \\
    CSI300 &  1.77 & \textbackslash & \textbackslash & 0.009 & \textbackslash & \textbackslash &  \textbackslash & 0.197 & 3.19 & \textbackslash \\
\end{longtable}
}

\section{Comparison between different quantformer strategies}
\begin{figure}[ht]
\centerline{\includegraphics[width=0.9\textwidth]{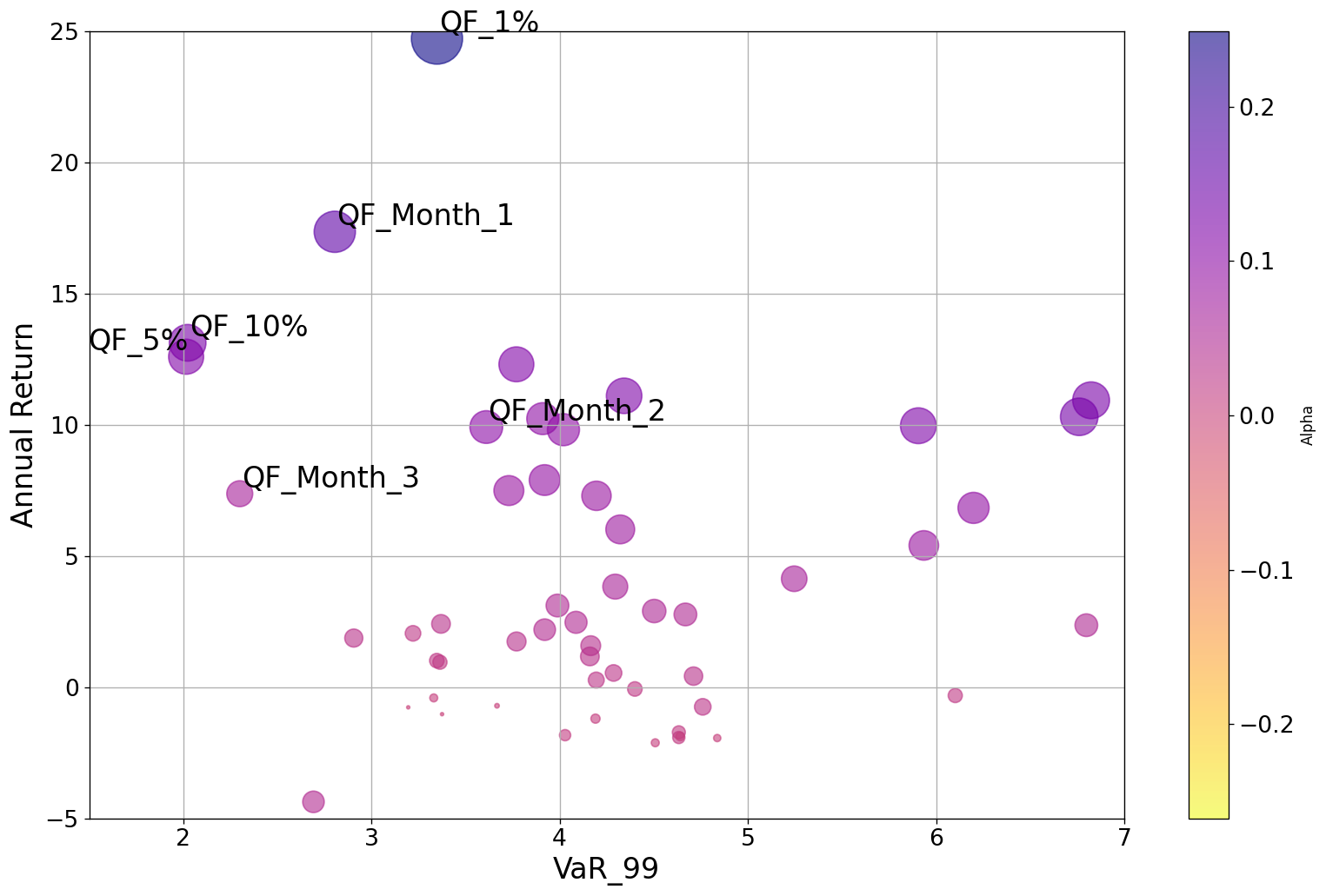}}
\renewcommand{\figurename}{Figure}
\caption{Comparison between factor performance and quantformer performance}
\label{scatter}
\end{figure}

In the provided scatterplot~\ref{scatter}, the $\mathrm{QF\_Month\_1}$ factor exhibits notable performance in terms of annual return. Compared to other factors, $\mathrm{QF\_Month\_1}$ offers better returns. Regarding risk, as measured by the \\ 
$\mathrm{Variance\_Covariance\_VaR\_99}$, the $\mathrm{QF\_Month\_1}$ factor's data points are positioned in a moderately high range, indicating a balanced risk profile. The size and color intensity of the $\mathrm{QF\_Month\_1}$ points, reflecting higher Alpha values, suggest a robust excess return or relative performance. Combining these observations, the $\mathrm{QF\_Month\_1}$ factor likely provides a compelling investment opportunity by delivering robust returns while maintaining a moderate risk level, an attractive proposition for strategies aiming to optimize the trade-off between risk and return. Overall, most of the factors based on quantformer perform better compared with the other 100 factors. 

\section{Symbol table}
\begin{table}[]
{\scalefont{0.90}
    \centering
    \begin{tabular}{l|l}
      Symbol  & Description \\
         \hline
        $A_{\mathrm{peak}}, A_{\mathrm{trough}}$ & peak value of the portfolio\\
       $\alpha$  & alpha ratio \\
       $\mathbf{b}$ & decision factor in the strategy\\
       $\theta$  & bias in the model \\
       $\beta_p$  & beta ratio \\
       $D_p$ & daily return \\
       $h, H$  & attention head \\
       $\chi^t_n$, $\tilde{\chi}^t_n$  & input feature matrix, normalized sequence \\
       $\hat{Y}^t_n$    & predicted output \\
       $\mathbb{E}[x]$  &  mean \\
       $\mathcal{G}^t$   &  set of the next timestamp profit \\
       $\mathrm{MD}$ & max drawdown \\
       $t$ & timestep \\ 
       $m$  & the number of timesteps before $t$ \\
       $\mathrm{MSELoss}$   & MSE loss \\
       $n$   &  the index of the stock \\
       $p^t$  & the close price of the stock \\
       $P$  & amount value of portfolio \\
       $\Phi$  & trading strategy \\
       $\Psi$  & the empirical quantile CDF \\
       $R_f$, $R_m$, $R_p$  & return of risk-free rate, market return, return of portfolio \\
       $r^t_n$ & profit rate \\
       $s^t_n$  &  stock \\
       $S^t$   & stock set \\
       $\mathrm{SR}$  &  Sharpe ratio \\
       $\mathrm{std}[x]$ & standard deviation \\
       $\mathrm{TR}$  &  turnover rate of the portfolio\\
       $v^t_n$  & turnover rate of the stock over the time period $t$\\
       $\mathrm{VaR}_{\alpha}$  &  value at risk at $\alpha$ level\\
       $w^t_n$   & weight of portfolio \\
       $x^t_n$, $\tilde{x}^t_n$   & row vector, normalized vector \\
       $y^t_n$, $\tilde{y}^t_n$   &  target label, sorted predicted output \\
       $\varrho$  & number of bins in $y^t_n$ \\
       $\varphi$ & percentage of stocks for each $\varrho$\\
       $\xi$ & middle percentage of $y^t_n$\\
       $\mathcal{L}$   &  loss of holding stock \\
       $\mathbf{F}$  &  multi-head attention output \\
       $\mathbf{K}$, $\mathbf{Q}$, $\mathbf{V}$  &  key, query, value in quantformer \\
       $\mathbf{W}$   & trainable weight matrix \\
       $\mathbf{Z}$  &  output layer \\
         \hline
    \end{tabular}
    \label{tab:my_label}
}
\end{table}

\end{document}